\documentclass[prd,superscriptaddress,twocolumn,nofootinbib,showpacs,preprintnumbers,floatfix]{revtex4}

\usepackage{mathrsfs}
\usepackage{amsfonts,amssymb,amsmath}
\usepackage{graphicx,epsfig}
\usepackage{multirow}

\begin{document}

\title{ The No-Scale Multiverse at the LHC } 

\thanks{
Submitted to the Fifth International Workshop DICE 2010:
Space, Time, Matter - Current Issues in
Quantum Mechanics and Beyond, September 13-17, 2010,
Castello Pasquini, Italy,
based on the invited talk by D.V.N.}
\author{Tianjun Li}

\affiliation{George P. and Cynthia W. Mitchell Institute for
Fundamental Physics, Texas A$\&$M University, College Station, TX 77843, USA }

\affiliation{Key Laboratory of Frontiers in Theoretical Physics,
Institute of Theoretical Physics, Chinese Academy of Sciences,
Beijing 100190, P. R. China }

\author{James A. Maxin}

\affiliation{George P. and Cynthia W. Mitchell Institute for
Fundamental Physics, Texas A$\&$M University, College Station, TX 77843, USA }

\author{Dimitri V. Nanopoulos}

\affiliation{George P. and Cynthia W. Mitchell Institute for
Fundamental Physics, Texas A$\&$M University, College Station, TX 77843, USA }

\affiliation{Astroparticle Physics Group, Houston Advanced Research Center (HARC),
Mitchell Campus, Woodlands, TX 77381, USA}

\affiliation{Academy of Athens, Division of Natural Sciences,
 28 Panepistimiou Avenue, Athens 10679, Greece }

\author{Joel W. Walker}

\affiliation{Department of Physics, Sam Houston State University,
Huntsville, TX 77341, USA }


\begin{abstract}
We present a contemporary perspective on the String Landscape and the Multiverse of
plausible string, M- and F-theory vacua, seeking to demonstrate a non-zero probability
for the existence of a universe matching our own observed physics within the solution ensemble.
We argue for the importance of No-Scale Supergravity as an essential common underpinning for the
spontaneous emergence of a cosmologically flat universe from the quantum ``nothingness''.
Our context is a highly detailed phenomenological probe of No-Scale ${\cal F}$-$SU(5)$,
a model representing the intersection of the $\cal{F}$-lipped $SU(5)\times U(1)_X$ Grand Unified Theory (GUT)
with extra TeV-Scale vector-like multiplets derived out of $\cal{F}$-theory, and the dynamics of No-Scale Supergravity.
The latter in turn imply a very restricted set of high energy boundary conditions.

We present a highly constrained ``Golden Point'' located
near $M_{1/2} = 455$~GeV and $\tan \beta = 15$ in the $\tan\beta-M_{1/2}$ plane,
and a highly non-trivial ``Golden Strip'' with $\tan\beta\simeq 15$,
$m_{\rm t} = 173.0$-$174.4$~GeV, $M_{1/2} = 455$-$481$~GeV, and $M_{\rm V} = 691$-$1020$~GeV,
which simultaneously satisfies all the known experimental constraints, featuring also
an imminently observable proton decay rate. We supplement this bottom-up phenomenological perspective
with a top-down theoretical analysis of the one-loop effective
Higgs potential.  A striking consonance is achieved via the dynamic determination of $\tan\beta$ and $M_{1/2}$
for fixed $Z$-boson mass at the local {\it minimum minimorum} of the potential, that being the secondary
minimization of the spontaneously broken electroweak Higgs vacuum $V_{\rm min}$.
By also indirectly determining the electroweak scale, we suggest
that this constitutes a complete resolution of the Standard Model gauge hierarchy problem.

Finally, we present the distinctive collider level signatures of No-Scale $\cal{F}$-$SU(5)$
for the $\sqrt{s} = 7$~TeV LHC, with $1~{\rm fb}^{-1}$ of integrated luminosity.
The characteristic feature is a light stop and gluino, both sparticles lighter than all other squarks,
generating a surplus of ultra-high multiplicity ($\ge 9$) hadronic jet events.
We propose modest alterations to the canonical background selection cut strategy which are expected to yield
significantly enhanced resolution of the characteristic ultra-high jet multiplicity $\cal{F}$-$SU(5)$ events,
while readily suppressing the contribution of all Standard Model processes, and allowing moreover a clear differentiation
from competing models of new physics, most notably minimal supergravity.
Detection by the LHC of the ultra-high jet signal would constitute a suggestive evocation of the intimately linked
stringy origins of $\cal{F}$-$SU(5)$, and could provide a glimpse into the underlying
structure of the fundamental string moduli, possibly even opening a darkened glass upon the hidden
workings of the No-Scale Multiverse. 
\end{abstract}


\preprint{ACT-05-11, MIFPA-11-11}

\maketitle


\section{Introduction}

The number of consistent, meta-stable vacua of string, M- or (predominantly)
F-theory flux compactifications which exhibit broadly plausible phenomenology,
including moduli stabilization and broken supersymmetry~\cite{Bousso:2000xa, Giddings:2001yu, Kachru:2003aw, Susskind:2003kw, Denef:2004ze, Denef:2004cf}, is popularly
estimated~\cite{Denef:2004dm,Denef:2007pq} to be of order $10^{500}$.  It is moreover
currently in vogue to suggest that degeneracy of common features
across these many ``universes'' might statistically isolate the physically realistic
universe from the vast ``landscape'', much as the entropy function coaxes the singular order
of macroscopic thermodynamics from the chaotic duplicity of the entangled quantum microstate.
We argue here though the counter point that we are not obliged {\it a priori} to live
in the likeliest of all universes, but only in one which is possible.  The existence
merely of a non-zero probability for our existence is sufficient.

We indulge for this effort the fanciful imagination that the ``Multiverse'' of string
vacua might exhibit some literal realization beyond our own physical sphere.
A single electron may be said to wander all histories through interfering apertures,
though its arrival is ultimately registered at a localized point on the target.
The journey to that destination is steered by the full dynamics of the theory, although
the isolated spontaneous solution reflects only faintly the richness of the solution ensemble.
Whether the Multiverse be reverie or reality, the conceptual superset of our own physics
which it embodies must certainly represent the interference of all navigable universal histories.

Surely many times afore has mankind's notion of the heavens expanded - the Earth dispatched
from its central pedestal in our solar system and the Sun rendered one among some hundred billion
stars of the Milky Way, itself reduced to one among some hundred billion galaxies.
Finally perhaps, we come to the completion of our Odyssey, by realizing that our
Universe is one of at least $10^{500}$ so possible, thus rendering the anthropic view
of our position in the Universe (environmental coincidences explained away by the availability of
$10^{11} \times 10^{11}$ solar systems) functionally equivalent to the anthropic view of the
origin of the Universe (coincidences in the form and content of physical laws explained away
by the availability, through dynamical phase transitions, of $10^{500}$ universes).
Nature's bounty has anyway invariably trumped our wildest anticipations, and though
frugal and equanimous in law, she has spared no extravagance or whimsy in its manifestation.

Our perspective should not be misconstrued, however, as complacent retreat into the tautology of
the weak anthropic principle.  It is indeed unassailable truism that an observed universe
must afford and sustain the life of the observer, including requisite constraints,
for example, on the cosmological constant~\cite{Weinberg:1987dv} and gauge hierarchy.
Our point of view, though, is sharply different; We should be able to
resolve the cosmological constant and gauge hierarchy problems through investigation
of the fundamental laws of our (or any single) Universe, its accidental and specific properties
notwithstanding, without resorting to the existence of observers.  In our view, the observer is the
output of, not the {\it raison d'\^etre} of, our Universe.  Thus, our attention is advance from this
base camp of our own physics, as unlikely an appointment as it may be, to the summit goal of the
master theory and symmetries which govern all possible universes.  In so seeking, our first
halting forage must be that of a concrete string model which can describe Nature locally.


\section{The Ensemble Multiverse}

The greatest mystery of Nature is the origin of the Universe itself.
Modern cosmology is relatively clear regarding the occurrence of a hot big bang,
and subsequent Planck, grand unification, cosmic inflation, lepto- and baryogenesis,
and electroweak epochs, followed by nucleosynthesis, radiation decoupling, and large
scale structure formation.  In particular, cosmic inflation can address the flatness and
monopole problems, explain homogeneity, and generate the fractional anisotropy of the
cosmic background radiation by quantum fluctuation of the inflaton
field~\cite{Guth:1980zm, Linde:1981mu, Albrecht:1982wi, Ellis:1982ws, Nanopoulos:1982bv}.
A key question though, is from whence the energy of the Universe arose.  Interestingly,
the gravitational field in an inflationary scenario can supply the required positive
mass-kinetic energy, since its potential energy becomes negative without bound,
allowing that the total energy could be exactly zero.

Perhaps the most striking revelation of the post-WMAP~\cite{Spergel:2003cb,Spergel:2006hy,Komatsu:2010fb} era is the decisive
determination that our Universe is indeed globally flat, {\it i.e.}~, with the net energy contributions from
baryonic matter $\simeq 4\%$, dark matter $\simeq 23\%$, and the cosmological constant (dark energy)
$\simeq 73\%$ finely balanced against the gravitational potential.  Not long ago, it was
possible to imagine the Universe, with all of its physics intact, hosting any arbitrary
mass-energy density, such that ``$k=+1$'' would represent a super-critical cosmology of
positive curvature, and ``$k=-1$'' the sub-critical case of negative curvature.  In
hindsight, this may come to seem as na\"{\i}ve as the notion of an empty infinite
Cartesian space.  The observed energy balance is highly suggestive of a fundamental symmetry
which protects the ``$k=0$'' critical solution, such that the physical constants of our
Universe may not be divorced from its net content.

This null energy condition licenses the speculative connection {\it ex nihilo} of our present
universe back to the primordial quantum fluctuation of an external system. Indeed, there is
nothing which quantum mechanics abhors more than nothingness.  This being the case, an extra
universe here or there might rightly be considered no extra trouble at all!  Specifically, it has
been suggested~\cite{Guth:1980zm, Linde:1981mu, Albrecht:1982wi, Steinhardt, Vilenkin:1983xq}
that the fluctuations of a dynamically evolved expanding universe might spontaneously produce
tunneling from a false vacuum into an adjacent (likely also false) meta-stable vacuum of lower
energy, driving a local inflationary phase, much as a crystal of ice or a bubble of steam may nucleate
and expand in a super-cooled or super-heated fluid during first order transition.  In this ``eternal
inflation'' scenario, such patches of space will volumetrically dominate by virtue of their
exponential expansion, recursively generating an infinite fractal array of causally disconnected
``Russian doll'' universes, nesting each within another, and each featuring
its own unique physical parameters and physical laws.

From just the specific location on the solution ``target'' where our own Universe landed, it may
be impossible to directly reconstruct the full theory.  Fundamentally, it may be impossible even in
principle to specify why our particular Universe is precisely as it is.  However, superstring theory and
its generalizations may yet present to us a loftier prize - the theory of the ensemble Multiverse.


\section{The Invariance of Flatness}

More important than any differences between various possible vacua are the
properties which might be invariant, protected by basic symmetries of the underlying mechanics.
We suppose that one such basic property must be cosmological flatness, so that
the seedling universe may transition dynamically across the boundary of its own creation,
maintaining a zero balance of some suitably defined energy function.  In practice, this
implies that gravity must be ubiquitous, its negative potential energy allowing for positive
mass and kinetic energy.  Within such a universe, quantum fluctuations may not again cause
isolated material objects to spring into existence, as their net energy must necessarily be
positive.  For the example of a particle with 
mass $m$ on the surface of the Earth, the ratio of gravitational
to mass energy is more than nine orders of magnitude too small
\begin{equation}
\left| -\frac{G_N M_E m}{R_E}  \right| \div  m c^2 \simeq 7 \times 10^{-10}~,~
\end{equation}
where $G_N$ is the gravitational constant, $c$ is the speed of light, 
and $M_E$ and $R_E$ are the mass and radius of the Earth, respectively.
Even in the limiting case of a Schwarzschild black hole of mass $M_{BH}$, 
a particle of mass $m$ at the horizon $R_{S}=2 G_N  M_{BH}/c^2$ has a gravitational potential 
which is only half of that required.
\begin{equation}
\left| -\frac{G_N M_{BH} m}{R_{S}} \right| = \frac{1}{2} m c^2
\end{equation}
It is important to note that while the energy density for the gravitational field is
surely negative in Newtonian mechanics, the global gravitational field energy is not
well defined in general relativity.  Unique prescriptions
for a stress-energy-momentum pseudotensor can be formulated though,
notably that of Landau and Lifshitz.  Any such stress-energy can,  however,
be made to vanish locally by general coordinate transformation, and it is 
not even entirely clear that the pseudotensor so applied is an appropriate general
relativistic object.  Given though that Newtonian gravity is the classical limit of 
general relativity, it is reasonable to suspect that the properly defined field
energy density will be likewise also negative, and that inflation is indeed consistent
with a correctly generalized notion of constant, zero total energy.

A universe would then be in this sense closed, an island unto itself, from the moment of
its inception from the quantum froth; Only a universe {\it in toto} might so
originate, emerging as a critically bound structure possessing profound density and
minute proportion, each as accorded against intrinsically defined scales (the analogous
Newton and Planck parameters and the propagation speed of massless fields),
and expanding or inflating henceforth and eternally.


\section{The Invariance of No-Scale SUGRA}

Inflation, driven by the scalar inflaton field is itself inherently a quantum field
theoretic subject.  However, there is tension between quantum mechanics and general
relativity.  Currently, superstring theory is the best candidate for quantum gravity.  The
five consistent ten dimensional superstring theories, namely heterotic $E_8 \times E_8$,
heterotic $SO(32)$, Type I, Type IIA, Type IIB, can be unified by various duality transformations
under an eleven-dimensional M-theory~\cite{Witten:1995ex}, and the twelve-dimensional 
F-theory can be considered as the strongly coupled formulation of the Type IIB
string theory with a varying axion-dilaton field~\cite{Vafa:1996xn}.  Self consistency of the string
(or M-, F-) algebra implies a ten (or eleven, twelve) dimensional master spacetime,
some elements of which -- six (or seven, eight) to match our observed four large dimensions --
may be compactified on a manifold (typically Calabi-Yau manifolds
or $G_2$ manifolds) which conserves a requisite portion
of supersymmetric charges.

The structure of the curvature within the extra dimensions dictates in no small measure
the particular phenomenology of the unfolded dimensions, secreting away the ``closet space''
to encode the symmetries of all gauged interactions.  The physical volume of the internal 
spatial manifold is directly related to the effective Planck scale and basic gauge coupling
strengths in the external space.  The compactification is in turn described by fundamental
moduli fields which must be stabilized, {\it i.e.}~, given suitable vacuum expectation values (VEVs).

The famous example of Kaluza and Klein prototypes the manner in which general covariance in five dimensions
is transformed to gravity plus Maxwell theory in four dimensions when the transverse fifth dimension
is cycled around a circle.  The connection of geometry to particle physics is perhaps nowhere
more intuitively clear than in the context of model building with $D6$-branes, where the
gauge structure and family replication are related directly to the brane stacking and intersection
multiplicities.  The Yukawa couplings and Higgs structure are in like manners also specified,
leading after radiative symmetry breaking of the chiral gauge sector to low energy masses
for the chiral fermions and broken gauge generators, each massless in the symmetric limit. 

From a top-down view, Supergravity (SUGRA) is an ubiquitous infrared limit of string theory,
and forms the starting point of any two-dimensional world sheet or D-dimensional target space action.
The mandatory localization of the Supersymmetry (SUSY) algebra, and thus the
momentum-energy (space-time translation) operators, leads to general coordinate
invariance of the action and an Einstein field theory limit.  Any available
flavor of Supergravity will not however suffice.  In general, extraneous
fine tuning is required to avoid a cosmological constant which scales like a dimensionally
suitable power of the Planck mass.  Neglecting even the question of whether such a universe might
be permitted to appear spontaneously, it would then be doomed to curl upon itself and collapse
within the order of the Planck time, for comparison about $10^{-43}$ seconds in our Universe.
Expansion and inflation appear to uniquely require properties which arise naturally only in the
No-Scale SUGRA formulation~\cite{Cremmer:1983bf,Ellis:1983sf, Ellis:1983ei, Ellis:1984bm, Lahanas:1986uc}.

SUSY is in this case broken while the vacuum energy density vanishes automatically at tree level due to
a suitable choice of the K\"ahler potential, the function which specifies the metric on superspace.
At the minimum of the null scalar potential, there are flat directions which leave the compactification
moduli VEVs undetermined by the classical equations of motion.  We thus receive without additional effort
an answer to the deep question of how these moduli are stabilized; They have been transformed into
dynamical variables which are to be determined by minimizing corrections to the scalar
potential at loop order.  In particular, the high energy gravitino mass $M_{3/2}$, and also the
proportionally equivalent universal gaugino mass $M_{1/2}$, will be established in this way.  Subsequently,
all gauge mediated SUSY breaking soft-terms will be dynamically evolved down from this boundary under the
renormalization group~\cite{Giudice:1998bp}, establishing in large measure the low energy phenomenology, and
solving also the Flavour Changing Neutral Current (FCNC) problem.  Since the moduli are fixed at a false local
minimum, phase transitions by quantum tunneling will naturally occur between discrete vacua.

The specific K\"ahler potential which we favor has been independently derived in both
weakly coupled heterotic string theory~\cite{Witten:1985xb} and the leading order compactification of
M-theory on $S^1/Z_2$~\cite{Li:1997sk}, and might be realized in F-theory models
as well~\cite{Beasley:2008dc,Beasley:2008kw, Donagi:2008ca, Donagi:2008kj}.  We conjecture, for the reasons given prior, that the No-Scale SUGRA construction
could pervade all universes in the String Landscape with reasonable flux vacua.  This being the case,
intelligent creatures elsewhere in the Multiverse, though separated from us by a bridge too far,
might reasonably so concur after parallel examination of their own physics.  Moreover, they might leverage via
this insight a deeper knowledge of the underlying Multiverse-invariant master theory, of which our known string,
M-, and F-theories may compose some coherently overlapping patch of the garment edge.  Perhaps we yet share
appreciation, across the cords which bind our 13.7 billion years to their corresponding blink of history,
for the common timeless principles under which we are but two isolated condensations upon two particular
vacuum solutions among the physical ensemble.


\section{An Archetype Model Universe}

Though we engage in this work lofty and speculative questions of natural philosophy,
we balance abstraction against the measured material underpinnings of concrete
phenomenological models with direct and specific connection to tested and testable particle physics.
If the suggestion is correct that eternal inflation and No-Scale SUGRA models with
string origins together describe what is in fact our Multiverse, then we must as a prerequisite
settle the issue of whether our own phenomenology can be produced out of such a construction.

In the context of Type II intersecting D-brane models, we have indeed found one realistic 
Pati-Salam model which might describe Nature as we observe
it~\cite{Cvetic:2004ui, Chen:2007px, Chen:2007zu}. 
If only the F-terms of three complex structure moduli are non-zero, we also
automatically have vanishing vacuum energy, and obtain a generalized 
No-Scale SUGRA. 
It seems to us that the string derived Grand Unified Theories (GUTs), and particularly the 
Flipped $SU(5)\times U(1)_X$ models~\cite{Barr:1981qv, Derendinger:1983aj, Antoniadis:1987dx}, are also candidate
realistic string models with promising predictions that can be tested at the Large
Hadron Collider (LHC), the Tevatron, and other future experiments.

Let us briefly review the minimal flipped
$SU(5)\times U(1)_X$ model~\cite{Barr:1981qv, Derendinger:1983aj, Antoniadis:1987dx}. 
The gauge group of the flipped $SU(5)$ model is
$SU(5)\times U(1)_{X}$, which can be embedded into $SO(10)$.
We define the generator $U(1)_{Y'}$ in $SU(5)$ as 
\begin{eqnarray} 
T_{\rm U(1)_{Y'}}={\rm diag} \left(-\frac{1}{3}, -\frac{1}{3}, -\frac{1}{3},
 \frac{1}{2},  \frac{1}{2} \right).
\label{u1yp}
\end{eqnarray}
The hypercharge is given by
\begin{eqnarray}
Q_{Y} = \frac{1}{5} \left( Q_{X}-Q_{Y'} \right).
\label{ycharge}
\end{eqnarray}
In addition, 
there are three families of SM fermions 
whose quantum numbers under the $SU(5)\times U(1)_{X}$ gauge group are
\begin{eqnarray}
F_i={\mathbf{(10, 1)}},~ {\bar f}_i={\mathbf{(\bar 5, -3)}},~
{\bar l}_i={\mathbf{(1, 5)}},
\label{smfermions}
\end{eqnarray}
where $i=1, 2, 3$. 

To break the GUT and electroweak gauge symmetries, we 
introduce two pairs of Higgs fields
\begin{eqnarray}
&H={\mathbf{(10, 1)}},~{\overline{H}}={\mathbf{({\overline{10}}, -1)}},& \\ \nonumber
&~h={\mathbf{(5, -2)}},~{\overline h}={\mathbf{({\bar {5}}, 2)}}.&
\label{Higgse1}
\end{eqnarray}
Interestingly, we can naturally solve the doublet-triplet splitting
 problem via the missing partner mechanism~\cite{Antoniadis:1987dx}, and then
the dimension five
proton decay from the colored Higgsino exchange can be
highly suppressed~\cite{Antoniadis:1987dx}.
The flipped $SU(5)\times U(1)_X$ models have been
constructed systematically in the free fermionic string 
constructions at Kac-Moody level one previously~\cite{Antoniadis:1987dx,Antoniadis:1987tv, Antoniadis:1988tt, Antoniadis:1989zy,Lopez:1992kg},
and in the  F-theory model building recently~\cite{Beasley:2008dc, Beasley:2008kw, Donagi:2008ca, Donagi:2008kj, Jiang:2009zza, Jiang:2009za}.

In the flipped $SU(5)\times U(1)_X$ models, there are two unification
scales: the $SU(3)_C\times SU(2)_L$ unification scale $M_{32}$ and
the $SU(5)\times U(1)_X$ unification scale $M_{\cal F}$.
To separate the $M_{32}$ and $M_{\cal F}$ scales
and obtain true string-scale gauge coupling unification in 
free fermionic string models~\cite{Jiang:2006hf, Lopez:1992kg} or
the decoupling scenario in F-theory models~\cite{Jiang:2009zza, Jiang:2009za},
we introduce vector-like particles which form complete
flipped $SU(5)\times U(1)_X$ multiplets.
In order to avoid the Landau pole
problem for the strong coupling constant, we can only introduce the
following two sets of vector-like particles around the TeV 
scale~\cite{Jiang:2006hf}
\begin{eqnarray}
&& Z1:  XF ={\mathbf{(10, 1)}}~,~
{\overline{XF}}={\mathbf{({\overline{10}}, -1)}}~;~\\
&& Z2: XF~,~{\overline{XF}}~,~Xl={\mathbf{(1, -5)}}~,~
{\overline{Xl}}={\mathbf{(1, 5)}}
~,~\,
\end{eqnarray}
where 
\begin{eqnarray}
{XF} ~\equiv~ (XQ,XD^c,XN^c)~,~~~{\overline{Xl}}_{\mathbf{(1, 5)}}\equiv XE^c ~.~\,
\end{eqnarray}
In the prior, $XQ$, $XD^c$, $XE^c$, $XN^c$ have the same quantum numbers as the
quark doublet, the right-handed down-type quark, charged lepton, and
neutrino, respectively. 
Such kind of the models have been constructed 
systematically in the F-theory model building locally and dubbed 
${\cal F}-SU(5)$ within that context~\cite{Jiang:2009zza, Jiang:2009za}.
In this paper, we only consider the flipped
$SU(5)\times U(1)_X$ models with 
$Z2$ set of vector-like particles.
The discussions for the models with 
$Z1$ set and heavy threshold corrections~\cite{Jiang:2009zza, Jiang:2009za}
are similar.

These models are moreover quite 
interesting from a phenomenological point of view~\cite{Jiang:2009zza, Jiang:2009za}. The predicted vector-like particles
can be observed at the Large Hadron Collider, and the partial lifetime for proton decay 
in the leading ${(e|\mu)}^{+} \pi^0 $ channels falls around 
$5 \times 10^{34}$ years, testable at the future 
Hyper-Kamiokande~\cite{Nakamura:2003hk} and
Deep Underground Science and Engineering 
Laboratory (DUSEL)~\cite{Raby:2008pd} experiments~\cite{Li:2010dp, Li:2010rz, Li:2009fq}.
The lightest CP-even Higgs boson mass can be increased~\cite{HLNT}, 
hybrid inflation can be naturally realized, and the 
correct cosmic primordial density fluctuations can be 
generated~\cite{Kyae:2005nv}.


\section{No-Scale Supergravity}

In the traditional framework, 
supersymmetry is broken in 
the hidden sector, and then its breaking effects are
mediated to the observable sector
via gravity or gauge interactions. In GUTs with
gravity mediated supersymmetry breaking, also known as the
minimal supergravity (mSUGRA) model, 
the supersymmetry breaking soft terms can be parameterized
by four universal parameters: the gaugino mass $M_{1/2}$,
scalar mass $M_0$, trilinear soft term $A$, and
the ratio of Higgs VEVs $\tan \beta$ at low energy,
plus the sign of the Higgs bilinear mass term $\mu$.
The $\mu$ term and its bilinear 
soft term $B_{\mu}$ are determined
by the $Z$-boson mass $M_Z$ and $\tan \beta$ after
the electroweak (EW) symmetry breaking.

To solve the cosmological constant
problem, No-Scale supergravity was proposed~\cite{Cremmer:1983bf,Ellis:1983sf, Ellis:1983ei, Ellis:1984bm, Lahanas:1986uc}. 
No-scale supergravity is defined as the subset of supergravity models
which satisfy the following three constraints~\cite{Cremmer:1983bf,Ellis:1983sf, Ellis:1983ei, Ellis:1984bm, Lahanas:1986uc}:
(i) The vacuum energy vanishes automatically due to the suitable
 K\"ahler potential; (ii) At the minimum of the scalar
potential, there are flat directions which leave the 
gravitino mass $M_{3/2}$ undetermined; (iii) The super-trace
quantity ${\rm Str} {\cal M}^2$ is zero at the minimum. Without this,
the large one-loop corrections would force $M_{3/2}$ to be either
zero or of Planck scale. A simple K\"ahler potential which
satisfies the first two conditions is
\begin{eqnarray} 
K &=& -3 {\rm ln}( T+\overline{T}-\sum_i \overline{\Phi}_i
\Phi_i)~,~
\label{NS-Kahler}
\end{eqnarray}
where $T$ is a modulus field and $\Phi_i$ are matter fields.
The third condition is model dependent and can always be satisfied in
principle~\cite{Ferrara:1994kg}.

The scalar fields in the above
K\"ahler potential parameterize the coset space
$SU(N_C+1, 1)/(SU(N_C+1)\times U(1))$, where $N_C$ is the number
of matter fields. Analogous structures appear in the 
$N\ge 5$ extended supergravity theories~\cite{Cremmer:1979up}, for example,
$N_C=4$ for $N=5$, which can be realized in the compactifications
of string theory~\cite{Witten:1985xb, Li:1997sk}. 
The non-compact structure of the symmetry
implies that the potential is not only constant but actually
identical to zero. In fact, one can easily check that
the scalar potential is automatically positive semi-definite,
and has a flat direction along the $T$ field. Interestingly,
for the simple K\"ahler potential in Equation~(\ref{NS-Kahler})
we obtain the simplest No-Scale boundary condition
$M_0=A=B_{\mu}=0$, while $M_{1/2}$ may be
non-zero at the unification scale,
allowing for low energy SUSY breaking. 

It is important to note that there exist several methods of generalizing
No-Scale supergravity, for instance, the previously mentioned Type II intersecting
D-brane models~\cite{Cvetic:2004ui, Chen:2007px, Chen:2007zu}, and
the compactifications of M-theory on $S^1/Z_2$ with next-to-leading order corrections,
where we have obtained a generalization employing modulus dominated SUSY 
breaking~\cite{Nilles:1997cm, Nilles:1998sx, Lukas:1997fg, Lukas:1998yy, Li:1998rn}.
Similarly, mirage mediation for the flux compactifications can be considered as
another form of generalized No-Scale supergravity~\cite{Choi:2004sx, Choi:2005ge}.
In this paper we concentrate on the simplest No-Scale supergravity,
reserving any such generalizations for future study.


\section{The Golden Point}

First, we would like to review the Golden Point of No-Scale and no-parameter 
${\cal F}$-$SU(5)$~\cite{Li:2010ws}.
In the No-Scale context, we impose $M_0 = A = B_\mu$ = 0 at the unification scale
$M_{\cal F}$, and allow distinct inputs for the single parameter $M_{1/2}(M_{\cal F})$ to 
translate under the RGEs to distinct low scale outputs of $B_\mu$ and the Higgs mass-squares $M^2_{H_u}$
and $M^2_{H_d}$.  This continues until the point of spontaneous breakdown of the electroweak symmetry at
$M^2_{H_u} + \mu^2 = 0$, at which point minimization of the broken potential establishes the
physical low energy values of $\mu$ and $\tan \beta$.
In practice however, this procedure is at odds with the existing
{\tt SuSpect 2.34} code~\cite{Djouadi:2002ze} base from which our primary routines have been adapted.
In order to impose the minimal possible refactoring, we have instead
opted for an inversion wherein $M_{1/2}$ and $\tan \beta$ float
as two effective degrees of freedom. Thus, we do not fix $B_\mu(M_{\cal F})$.
We take $\mu >0$ as suggested by the results of $g_{\mu}-2$ for the muon,
and assume as in prior work that the masses for the vector-like particles are 
universal at 1 TeV.

The relic LSP neutralino density, WIMP-nucleon direct 
detection cross sections  and 
photon-photon annihilation cross sections
are computed with {\tt MicrOMEGAs 2.1}~\cite{Belanger:2008sj} 
wherein the revised {\tt SuSpect} RGEs have also implemented.
We use a top quark mass of $m_{t}$ = 173.1 GeV~\cite{:2009ec} and 
employ the following experimental constraints:
(1) The WMAP 7-year measurements of 
the cold dark matter density~\cite{Spergel:2003cb,Spergel:2006hy,Komatsu:2010fb}, 
0.1088 $\leq \Omega_{\chi} \leq$ 0.1158. We allow $\Omega_{\chi}$ to be
larger than the upper bound due to a possible $\cal{O}$(10) 
dilution factor~\cite{Mavromatos:2009pm}
and to be smaller than the lower bound due to multicomponent
dark matter. (2) The experimental limits on 
the FCNC process, $b \rightarrow s\gamma$. 
We use the limits 
$2.86 \times 10^{-4} \leq Br(b \rightarrow s\gamma) 
\leq 4.18 \times 10^{-4}$~\cite{Barberio:2007cr, Misiak:2006zs}.
(3) The anomalous magnetic moment of the muon, $g_{\mu} - 2$. 
We use the $2\sigma$ level boundaries, 
$11 \times 10^{-10} < \Delta a_{\mu} < 44 \times 10^{-10}$~\cite{Bennett:2004pv}. 
(4) The process $B_{s}^{0} \rightarrow \mu^+ \mu^-$ where we take the upper bound to be
 $Br(B_{s}^{0} \rightarrow \mu^{+}\mu^{-}) < 5.8 \times 10^{-8}$~\cite{:2007kv}. 
(5) The LEP limit on the lightest CP-even Higgs boson 
mass, $m_{h} \geq 114$ GeV~\cite{Barate:2003sz,Yao:2006px}.

\begin{figure}[ht]
	\centering
		\includegraphics[width=0.4\textwidth]{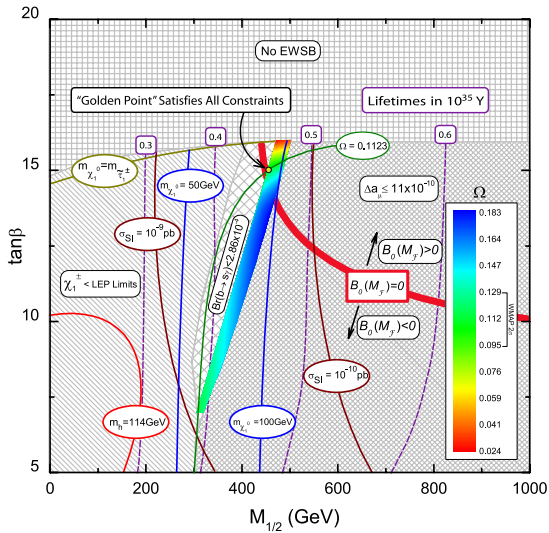}
		\caption{Viable parameter space in the $\tan\beta-M_{1/2}$ plane.
The ``Golden Point'' is annotated. 
The thin, dark green line denotes the WMAP 7-year central value $\Omega_{\chi} = 0.1123$. The dashed purple contours
label $p \!\rightarrow\! {(e\vert\mu)}^{\!+}\! \pi^0$ proton lifetime 
predictions, in units of $10^{35}$ years.}
	\label{fig:NoScale_M12_vs_tanb}
\end{figure}

In the $\tan\beta-M_{1/2}$ plane,
$B_\mu(M_{\cal F})$ is then calculated along
with the low energy supersymmetric particle
 spectrum and checks on various experimental constraints.
The subspace corresponding to a No-Scale model is clearly
then a one dimensional slice of this manifold, as demonstrated in Figure~\ref{fig:NoScale_M12_vs_tanb}.
It is quite remarkable that
the $B_\mu(M_{\cal F}) = 0$ contour so established runs sufficiently perpendicular to the WMAP
strip that the point of intersection effectively absorbs our final degree
of freedom, creating what we have labeled as a No-Parameter Model.  It is truly
extraordinary however that this intersection occurs exactly at the centrally
preferred relic density, that being our strongest experimental constraint.  We emphasize
again that there did not have to be an experimentally viable $B_\mu(M_{\cal F}) = 0$
solution, and that the consistent realization of this scenario depended crucially
on several uniquely identifying characteristics of the underlying proposal.  Specifically
again, it appears that the No-Scale condition comes into its own only when applied at
near the Planck mass, and that this is naturally identified as
the point of the final ${\cal F}$-$SU(5)$ unification, which is naturally extended and
decoupled from the primary GUT scale only via the modification to the RGEs from the
TeV scale ${\cal F}$-theory vector-like multiplet content.  The union of our
top-down model based constraints with the bottom-up experimental data exhausts
the available freedom of parameterization in a uniquely consistent and predictive manner,
phenomenologically defining a truly Golden Point
near $M_{1/2} = 455$ GeV and $\tan\beta$ = 15 GeV. 


\begin{table}[ht]
  \small
	\centering
	\caption{Spectrum (in GeV) for the Golden Point of Figure~\ref{fig:NoScale_M12_vs_tanb}. 
Here, $\Omega_{\chi}$ = 0.1123, $\sigma_{SI} = 1.9 \times 10^{-10}$ pb, and
$\left\langle \sigma v \right\rangle_{\gamma\gamma} = 1.7 \times 10^{-28} ~cm^{3}/s$.
The central prediction for the $p \!\rightarrow\! {(e\vert\mu)}^{\!+}\! \pi^0$ 
proton lifetime is $4.6 \times 10^{34}$ years.}
		\begin{tabular}{|c|c||c|c||c|c||c|c||c|c||c|c|} \hline		
    $\widetilde{\chi}_{1}^{0}$&$95$&$\widetilde{\chi}_{1}^{\pm}$&$185$&$\widetilde{e}_{R}$&$150$&$\widetilde{t}_{1}$&$489$&$\widetilde{u}_{R}$&$951$&$m_{h}$&$120.1$\\ \hline
    $\widetilde{\chi}_{2}^{0}$&$185$&$\widetilde{\chi}_{2}^{\pm}$&$826$&$\widetilde{e}_{L}$&$507$&$\widetilde{t}_{2}$&$909$&$\widetilde{u}_{L}$&$1036$&$m_{A,H}$&$920$\\ \hline
    
    $\widetilde{\chi}_{3}^{0}$&$821$&$\widetilde{\nu}_{e/\mu}$&$501$&$\widetilde{\tau}_{1}$&$104$&$\widetilde{b}_{1}$&$859$&$\widetilde{d}_{R}$&$992$&$m_{H^{\pm}}$&$925$\\ \hline
    $\widetilde{\chi}_{4}^{0}$&$824$&$\widetilde{\nu}_{\tau}$&$493$&$\widetilde{\tau}_{2}$&$501$&$\widetilde{b}_{2}$&$967$&$\widetilde{d}_{L}$&$1039$&$\widetilde{g}$&$620$\\ \hline
		\end{tabular}
		\label{tab:masses}
\end{table}

\begin{figure}[ht]
	\centering
		\includegraphics[width=0.4\textwidth]{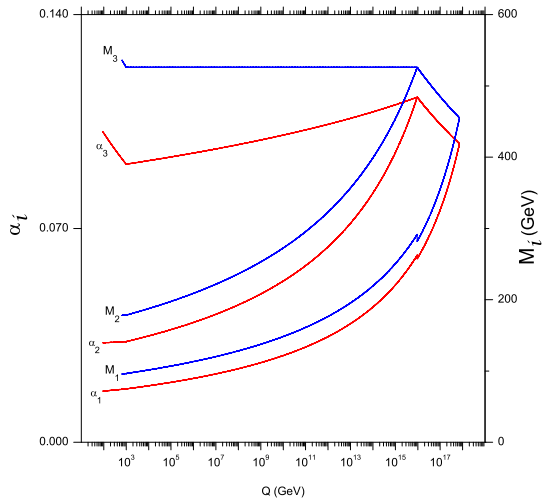}
		\caption{ RGE Running of the SM gauge couplings and gaugino masses from 
the EW scale to the unification scale $M_{\cal F}$.  There is a discontinuity in the running
at $M_{32}$ due to remixing of the hypercharge with the residual Abelian phase from the
breaking of SU(5)}
	\label{fig:NoScale_alpha}
\end{figure}

Our Golden point features $M_{1/2} = 455.3$ GeV, $\tan \beta = 15.02$,
and is in full compliance with the WMAP 7-year 
results, with $\Omega_{\chi}$ = 0.1123.
It also satisfies the CDMS~II~\cite{Ahmed:2008eu},
Xenon~100~\cite{Aprile:2010um}, and
FERMI-LAT space telescope constraints~\cite{Abdo:2010dk}, with $\sigma_{SI} = 1.9 \times 10^{-10}$ pb and 
$\left\langle \sigma v \right\rangle_{\gamma\gamma} = 1.7 \times 10^{-28} ~cm^{3}/s$. 
The proton lifetime is about $4.6\times 10^{34}$ years, which
is well within reach of the upcoming Hyper-Kamiokande~\cite{Nakamura:2003hk}
and DUSEL~\cite{Raby:2008pd} experiments. Inspecting the supersymmetric particle and Higgs spectrum
for the Golden Point of Table~\ref{tab:masses} reveals that the additional 
contribution of the 1 TeV vector-like particles lowers the gluino mass quite 
dramatically. The gluino mass $M_{3}$ runs flat from the $M_{32}$ unification 
scale to 1 TeV as shown in Figure~\ref{fig:NoScale_alpha}, though, 
due to supersymmetric radiative corrections, the physical gluino mass at the EW scale 
is larger than $M_{3}$ at the $M_{32}$ scale. This is true 
for the full parameter space. For our data point, 
the LSP neutralino is 99.8\% Bino. Similarly to the mSUGRA picture,
our this point is in
the stau-neutralino coannihilation region, but the gluino is lighter
than the squarks in our models.

\begin{figure}[ht]
	\centering
		\includegraphics[width=0.4\textwidth]{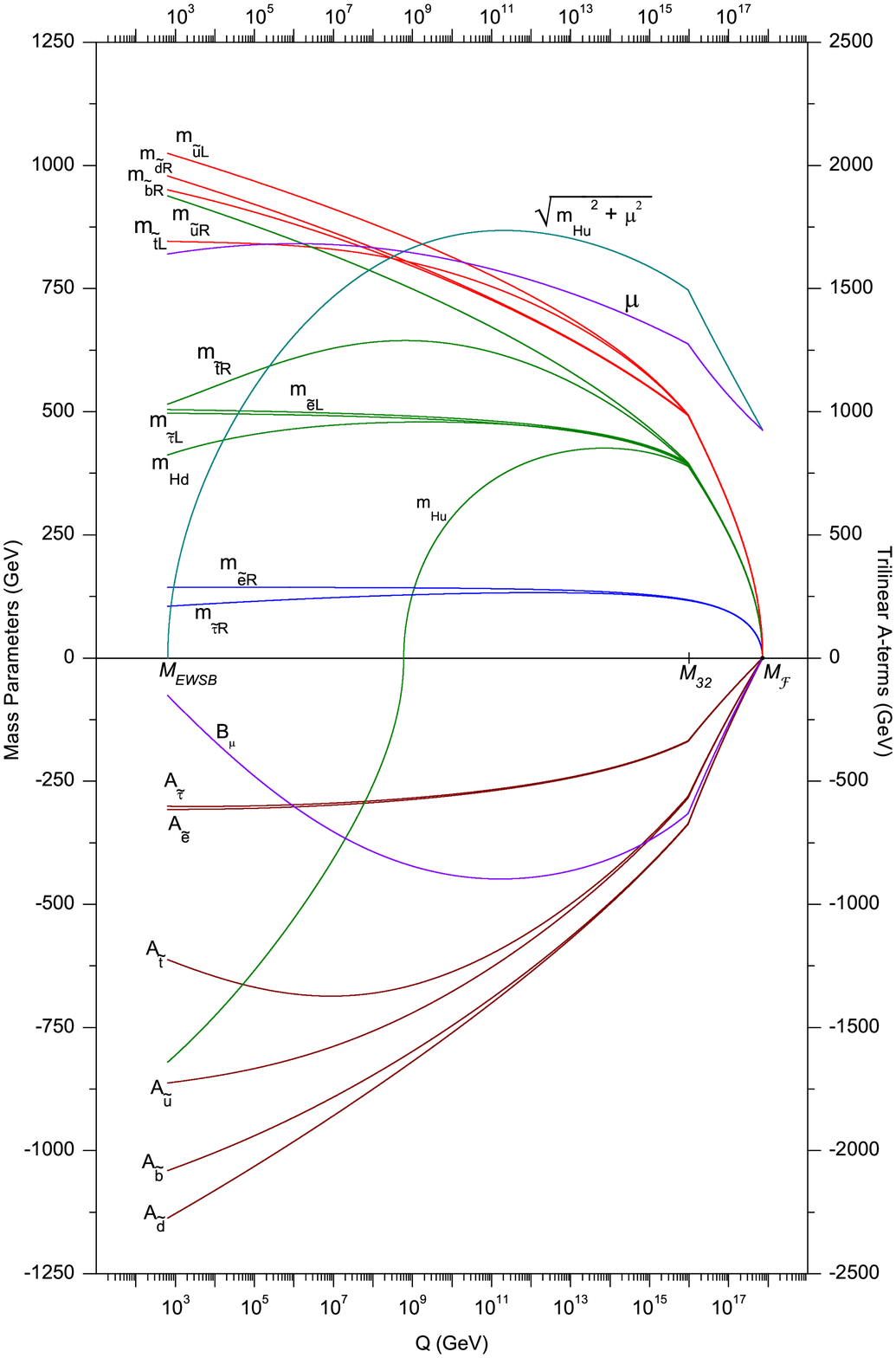}
		\caption{RGE Running of the $\mu$ term and SUSY breaking
soft terms from the EW scale to the unification scale $M_{\cal F}$.}
	\label{fig:NoScale_parameters}
\end{figure}

We plot gauge coupling and gaugino mass unification for the Golden Point 
in Figure~\ref{fig:NoScale_alpha}.
The figure explicitly demonstrates the two-step unification of flipped $SU(5)\times U(1)_X$.
In addition, we present the RGE running for the $\mu$ term,
the SUSY breaking scalar masses, trilinear A-terms, and bilinear $B_{\mu}$ term
in Figure~\ref{fig:NoScale_parameters}.
Note in particular that the EW symmetry breaking occurs when $H_{u}^{2} + \mu^{2}$ goes negative.

\section{ The Golden Strip}

\begin{figure}[ht]
	\centering
	\includegraphics[width=0.4\textwidth]{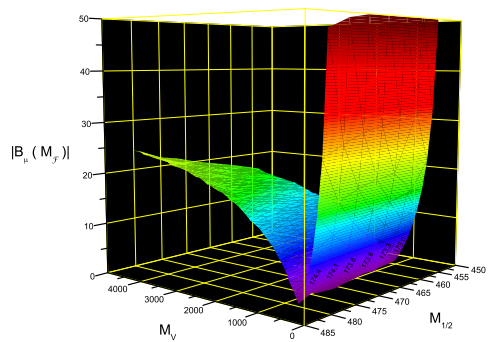}
	\caption{
The $B_\mu = 0$ target for
variations in $(M_{1/2},M_{\rm V})$, with $\tan \beta = 15$.
The specific $m_{\rm t}$ which is required to minimize
$\vert B_\mu(M_{\cal{F}}) \vert$ is annotated along the solution string.
	}
        \label{fig:bvalley}
\end{figure}

Second, we shall review the Golden Strip of correlated top quark, gaugino, and vectorlike mass 
in No-Scale, no-parameter ${\cal F}$-$SU(5)$~\cite{Li:2010mi}.
From the above discussions, only a small portion of viable
parameter space is consistent with
the $B_{\mu}(M_{\cal F}) = 0$ condition, which thus
constitutes a strong constraint. Since the boundary value of the universal gaugino mass $M_{1/2}$, and
even the unification scale $M_{\cal F} \simeq 7.5 \times 10^{17}$ GeV itself, are
established by the low energy experiments via
RGE running, we are not left with
any surviving scale parameters in the present model. The floor of the ``valley gorge" in
Figure~{\ref{fig:bvalley}} represents accord with the $B_\mu = 0$ target for
variations in $(M_{1/2},M_{\rm V})$. We fix $\tan \beta = 15$,
as appears to be rather generically required in No-Scale $\cal{F}$-$SU(5)$
to realize radiative EWSB and match the observed CDM density.

\begin{figure}[ht]
	\centering
	\includegraphics[width=0.4\textwidth]{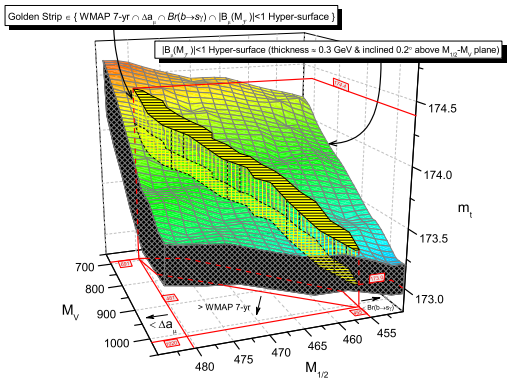}
	\caption{
With $\tan \beta \simeq 15$ fixed by WMAP-7, the residual parameter volume is three dimensional
in $(M_{1/2},M_{\rm V},m_{\rm t})$, with the $\vert B_\mu(M_{\cal{F}}) \vert \leq 1$ (slightly thickened)
surface forming a shallow $(0.2^\circ)$ incline above the $(M_{1/2},M_{\rm V})$ plane.
	}
        \label{fig:goldenstrip}
\end{figure}

\begin{figure}[ht]
	\centering
	\includegraphics[width=0.4\textwidth]{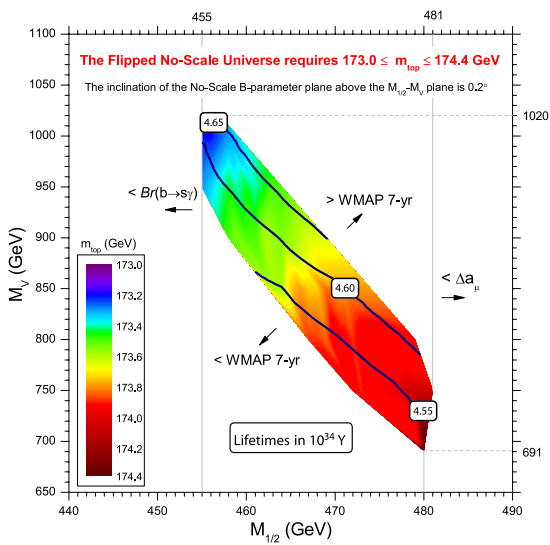}
	\caption{
A flattened presentation of the the $(M_{12},M_{\rm V})$ plane depicted in Figure~\ref{fig:goldenstrip}.
The overlayed blue contours
mark the $p \!\rightarrow\! {(e\vert\mu)}^{\!+}\! \pi^0$ proton lifetime
prediction, in units of $10^{34}$ years.
	}
        \label{fig:goldenstrip2}
\end{figure}

We have allowed for uncertainty
in the most sensitive experimental input, the top quark mass,
by effectively redefining $m_{\rm t}$ as an independent free parameter.
Lesser sensitivities to uncertainty in $(\alpha_{\rm s},M_{\rm Z})$ are included
in the $\pm 1$~GeV deviation from strict adherence to $B_\mu = 0$.
We have established that there is a two dimensional sheet (of some marginal thickness
to recognize the mentioned uncertainty) defining $\vert B_\mu(M_{\cal{F}}) \vert \leq 1$
for each point in the three dimensional ($M_{1/2}$, $M_{\rm V}$, $m_{\rm t}$) volume,
as shown in Figures~(\ref{fig:goldenstrip},\ref{fig:goldenstrip2}).
This sheet is inclined in the region of interest at the very shallow angle of
$0.2^\circ$ to the ($M_{1/2}$,$M_{\rm V}$) plane, such that $m_{\rm t}$ is largely decoupled
from variation in the plane.

The $(g-2)_\mu$ and $b\rightarrow s\gamma$ constraints
vary most strongly with $M_{1/2}$.
The two considered effects are each at their lower limits at the boundary,
but they exert pressure in opposing directions on $M_{1/2}$ due to the fact that the
leading gaugino and squark contributions to $Br(b\rightarrow s\gamma)$ enter with an opposing sign to the SM
term and Higgs contribution. For the non-SM contribution to $\Delta a_\mu$, the effect is additive,
and establishes an upper mass limit on $M_{1/2}$.
Incidentally, the same experiment forms the central rationale for the adoption of
$\rm{sign}(\mu) > 0$, such that appropriate interference terms between SM and SUSY
contributions are realized.
Conversely, the requirement that SUSY contributions to $Br(b\rightarrow s \gamma)$ not be overly large,
undoing the SM effect, requires a sufficiently large, {\it i.e.}~lower bounded, $M_{1/2}$.
The WMAP-7 CDM measurement, by contrast, exhibits a fairly strong correlation with both $(M_{1/2},M_{\rm V})$, cross-cutting the $M_{1/2}$ bound, and confining the vectorlike mass to 691-1020 GeV. We note that the mixing of the SM fermions and vector particles may give additional contributions to $Br(b\rightarrow s\gamma)$ and $\Delta a_\mu$, but we do not consider them here.

The intersection of these three key constraints with the
$\vert B_\mu(M_{\cal{F}}) \vert \leq 1$ surface, as depicted in
Figures~(\ref{fig:goldenstrip},\ref{fig:goldenstrip2}), defines the
``Golden Strip'' of No-Scale $\cal{F}$-$SU(5)$.
All of the prior is accomplished with no reference
to the experimental top quark mass, redefined here as a {\it free} input.
However, the extremely shallow angle of inclination ($0.2^\circ$) of
the $\vert B_\mu(M_{\cal{F}}) \vert \leq 1$ sheet
above the $(M_{1/2},M_{\rm V})$ plane and into the $m_{\rm t}$ axis
causes the Golden Strip to {\it imply} an exceedingly narrow
range of compatibility for $m_{\rm t}$, between $173.0$-$174.4$~GeV,
in perfect alignment with the physically observed value
of $m_{\rm t}$ = $173.1\pm 1.3$ GeV~\cite{:2009ec}.

Within the Golden Strip, we select the benchmark 
point of Table~\ref{tab:masses_2}. The Golden Strip is further consistent with the CDMS~II~\cite{Ahmed:2008eu} and 
Xenon~100~\cite{Aprile:2010um}
upper limits, with the spin-independent cross section extending from 
$\sigma_{SI} = 1.3$-$1.9 \times 10^{-10}$ pb.
Likewise, the allowed region satisfies the 
Fermi-LAT space telescope constraints~\cite{Abdo:2010dk}, with the 
photon-photon annihilation cross section 
$\left\langle \sigma v \right\rangle_{\gamma\gamma}$ ranging 
from $\left\langle \sigma v \right\rangle_{\gamma\gamma} = 1.5$-$1.7 \times 10^{-28} ~cm^{3}/s$.

\begin{table}[htb]
  \small
	\centering
	\caption{Spectrum (in GeV) for the benchmark point. 
Here, $M_{1/2}$ = 464 GeV, $M_{V}$ = 850 GeV, $m_{t}$ = 173.6 GeV, $\Omega_{\chi}$ = 0.112, $\sigma_{SI} = 1.7 \times 10^{-10}$ pb, and
$\left\langle \sigma v \right\rangle_{\gamma\gamma} = 1.7 \times 10^{-28} ~cm^{3}/s$.
The central prediction for the $p \!\rightarrow\! {(e\vert\mu)}^{\!+}\! \pi^0$ 
proton lifetime is $4.6 \times 10^{34}$ years. The lightest neutralino is 99.8\% Bino.}
		\begin{tabular}{|c|c||c|c||c|c||c|c||c|c||c|c|} \hline		
    $\widetilde{\chi}_{1}^{0}$&$96$&$\widetilde{\chi}_{1}^{\pm}$&$187$&$\widetilde{e}_{R}$&$153$&$\widetilde{t}_{1}$&$499$&$\widetilde{u}_{R}$&$975$&$m_{h}$&$120.6$\\ \hline
    $\widetilde{\chi}_{2}^{0}$&$187$&$\widetilde{\chi}_{2}^{\pm}$&$849$&$\widetilde{e}_{L}$&$519$&$\widetilde{t}_{2}$&$929$&$\widetilde{u}_{L}$&$1062$&$m_{A,H}$&$946$\\ \hline
     $\widetilde{\chi}_{3}^{0}$&$845$&$\widetilde{\nu}_{e/\mu}$&$513$&$\widetilde{\tau}_{1}$&$105$&$\widetilde{b}_{1}$&$880$&$\widetilde{d}_{R}$&$1018$&$m_{H^{\pm}}$&$948$\\ \hline
    $\widetilde{\chi}_{4}^{0}$&$848$&$\widetilde{\nu}_{\tau}$&$506$&$\widetilde{\tau}_{2}$&$514$&$\widetilde{b}_{2}$&$992$&$\widetilde{d}_{L}$&$1065$&$\widetilde{g}$&$629$\\ \hline
		\end{tabular}
		\label{tab:masses_2}
\end{table}


\section{ The Super No-Scale Mechanism }

In the following sections, we would like to review our study on 
Super No-Scale ${\cal F}$-$SU(5)$~\cite{Li:2010uu, Li:2011dw}.
The single relevant modulus field in the simplest 
string No-Scale supergravity is the K\"ahler
modulus $T$, a characteristic of the Calabi-Yau manifold,
the dilaton coupling being irrelevant.
The F-term of $T$ generates the gravitino mass $M_{3/2}$, 
which is proportionally equivalent to $M_{1/2}$.
Exploiting the simplest No-Scale boundary condition at $M_{\cal F}$ and 
running from high energy to low energy under the RGEs,
there can be a secondary minimization, or {\it minimum minimorum}, of the minimum of the
Higgs potential $V_{\rm min}$ for the EWSB vacuum.
Since $V_{\rm min}$ depends on $M_{1/2}$, the gaugino mass $M_{1/2}$ is consequently 
dynamically determined by the equation $dV_{\rm min}/dM_{1/2}=0$,
aptly referred to as the ``Super No-Scale'' mechanism~\cite{Li:2010uu, Li:2011dw}.

It could easily have been that in consideration of the above technique, there were: A) too few undetermined parameters,
with the $B_{\mu}=0$ condition forming an incompatible over-constraint, and thus demonstrably false, or B) so many
undetermined parameters that the dynamic determination possessed many distinct solutions, or was so far separated
from experiment that it could not possibly be demonstrated to be true.  The actual state of affairs is much
more propitious, being specifically as follows. The three parameters $M_0,A,B_{\mu}$ are once again identically zero at the
boundary because of the defining K\"ahler potential, and are thus known at all other scales as well by the RGEs.  The
minimization of the Higgs scalar potential with respect to the neutral elements of both SUSY Higgs doublets gives two
conditions, the first of which fixes the magnitude of $\mu$.  The second condition, which would traditionally be used
to fix $B_{\mu}$, instead here enforces a consistency relationship on the remaining parameters, being that
$B_{\mu}$ is already constrained.

In general, the $B_{\mu} = 0$ condition gives a hypersurface of solutions cut out from a very large parameter space.
If we lock all but one parameter, it will give the final value.  If we take a slice of two dimensional space, as has been 
described, it will give a relation between two parameters for all others fixed.
In a three-dimensional view with $B_{\mu}$ on the vertical axis, this
curve is the ``flat direction'' line along the bottom of the trench of $B_{\mu}=0$ solutions.  In general, we
must vary at least two parameters rather than just one in isolation, in order that their mutual compensation may transport
the solution along this curve.  The most natural first choice is in some sense the pair of prominent unknown inputs 
$M_{1/2}$ and $\tan \beta$, as was demonstrated in Ref.~\cite{Li:2010uu, Li:2011dw}.

Having come to this point, it is by no means guaranteed that the potential
will form a stable minimum.  It must be emphasized that the $B_{\mu}=0$ No-Scale
boundary condition is the central agent affording this determination, as it is the extraction of the parameterized
parabolic curve of solutions in the two compensating variables which allows for a localized, bound nadir point to be
isolated by the Super No-Scale condition, dynamically determining {\it both} parameters.  The background surface of
$V_{\rm min}$ for the full parameter space outside the viable $B_{\mu}=0$ subset is, in contrast, a steadily inclined
and uninteresting function. 
Although we have remarked that $M_{1/2}$
and $\tan \beta$ have no {\it directly} established experimental values, they are severely indirectly constrained by
phenomenology in the context of this model~\cite{Li:2010ws,Li:2010mi}.  It is highly non-trivial that there should be
accord between the top-down and bottom-up perspectives, but this is 
indeed precisely what has been observed~\cite{Li:2010uu, Li:2011dw}. 

\begin{figure}[htf]
        \centering
        \includegraphics[width=0.4\textwidth]{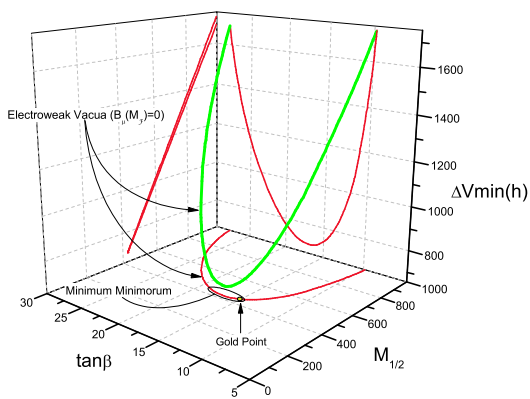}
        \caption{
The {\it minimum} $V_{\rm min}$ of the Higgs effective potential (green curve, GeV)
is plotted as a function of $M_{1/2}$ (GeV) and $\tan \beta$, 
emphasizing proximity of the ``Golden Point'' of Ref.~\cite{Li:2010ws}
to the dynamic region of the $V_{\rm min}$ {\it minimorum}.
        }
        \label{fig:vmin}
\end{figure}

\begin{figure}[htf]
        \centering
        \includegraphics[width=0.4\textwidth]{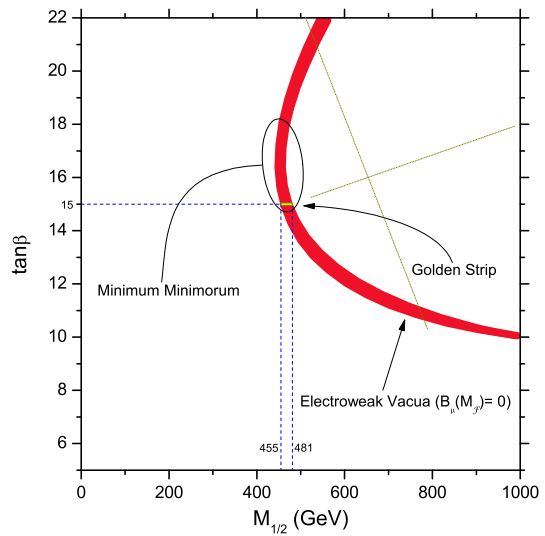}
        \caption{
The projection onto the ($M_{1/2}$,tan$\beta$) plane of Figure~\ref{fig:vmin} is further
detailed, expanding to span the boundary cases of
the Ref.~\cite{Li:2010mi} ``Golden Strip''.  The symmetry axis of the $B_{\mu} = 0$
parabola is rotated slightly above the $M_{1/2}$ axis.
        }
        \label{fig:vmin2}
\end{figure}


\section{The Minimum Minimorum of the Higgs Potential: Fixing $M_Z$}

We employ an effective
Higgs potential in the 't Hooft-Landau gauge and the $\overline{\rm DR}$
scheme, given summing the following neutral tree $(V_0)$ and one loop $(V_1)$ terms.
\begin{eqnarray}
V_0&=& (\mu^2 + m_{H_u}^2) (H_u^0)^2 + (\mu^2 + m_{H_d}^2) (H_d^0)^2
\nonumber \\ &&
-2 \mu B_{\mu} H_u^0 H_d^0 + {\frac{g_2^2 + g_Y^2}{8}} \left[(H_u^0)^2-(H_d^0)^2\right]^2 \nonumber
\label{higgs_v0} \\
V_1 &=&  \sum_i {\frac{n_i}{64\pi^2}} m_i^4(\phi)
\left( {\rm ln}{\frac{m_i^2(\phi)}{Q^2}} -{\frac{3}{2}} \right)
\label{higgs_v1}
\end{eqnarray}
Above, $m_{H_u}^2$ and $m_{H_d}^2$ are the SUSY
breaking soft masses of the Higgs fields $H_u$ and $H_d$,
$g_2$ and $g_Y$ are the gauge couplings of $SU(2)_L$ and
$U(1)_Y$, $n_i$ and $m_i^2(\phi)$ are the degree of freedom and mass
for $\phi_i$, and $Q$ is the renormalization scale. 
We include the complete Minimal Supersymmetric Standard Model
 (MSSM) contributions to one loop,
following Ref.~\cite{Martin:2002iu}, although the result is phenomenologically
identical accounting only the leading top and partner stop terms.
Since the minimum of the electroweak (EW) Higgs potential
$V_{\rm min}$ depends implicitly on $M_{3/2}$,  the gravitino mass is
determined by $dV_{\rm min}/dM_{3/2}=0$.
Being that $M_{1/2}$ is proportional to $M_{3/2}$,
it is equivalent to employ $M_{1/2}$
directly as our modulus parameter,
from which all other SUSY breaking soft terms here derive.
In our numerical results in the figures, we shall designate differences in
the fourth-root of the effective Higgs potential as $\Delta V_{min}(h)\equiv V_{\rm eff}^{1/4}$,
measured in units of GeV relative to an arbitrary overall zero-offset.

Factors explicit within the potential are obtained from 
our customized extension of the {\tt SuSpect 2.34}~\cite{Djouadi:2002ze}
codebase, including a self-consistency
assessment~\cite{Li:2010ws} on $B_\mu = 0$.
We apply two-loop RGE running for the SM gauge couplings,
and one-loop running for the SM fermion Yukawa couplings, 
$\mu$ term and soft terms.

Studying $V_{\rm min}$ generically in the $(M_{1/2}, \tan \beta)$ plane, no point of secondary
minimization is readily apparent in the strong, roughly linear, downward trend with
respect to $M_{1/2}$ over the region of interest.  However, the majority of the plane
is not in physical communication with our model, disrespecting the fundamental
$B_\mu = 0$ condition.  Isolating only the compliant contour within this surface,
{\it mirabile dictu}, a parabola is traced with nadir alighting gentle upon
our original Golden Point, as in Figure~(\ref{fig:vmin}).
Restoring parameterization freedom to $(M_{\rm V},m_{\rm t})$, we may scan across the corresponding
Golden Point of each nearby universe variant, reconstructing in their union the previously
advertised Golden Strip, as in Figure~(\ref{fig:vmin2}).  Notably, the theoretical
restriction on $\tan\beta$ remains stable against variation in these parameters,
exactly as its experimental counterpart. 
We find it quite extraordinary that the phenomenologically preferred region
rests precisely at the curve's locus of symmetric inflection.
Note in particular that it is the selection of the parabolic $B_\mu=0$ contour
out of the otherwise uninteresting $V_{\rm min}(M_{1/2},\tan \beta)$ inclined surface
which allows a clear {\it minimum minimorum} to be established.
We reiterate that consistency of the dynamically positioned
$M_{1/2}$ and $\tan\beta$ with the Golden Strip,
implies consistency with all current experimental data.

A strongly linear relationship is observed between the SUSY
and EWSB scales with $M_{\rm EWSB} \simeq 1.44~M_{1/2}$, such that a corresponding 
parabolic curve may be visualized.
There is a charged stau LSP for $\tan\beta$ from 16 to 22,
and we connect points with correct EWSB smoothly on the plot in this region.
If $\tan\beta$ is larger than 22, the stau is moreover
tachyonic, so properly we must restrict all analysis to $\tan\beta \leq 22$.


\section{The Gauge Hierarchy Problem}

Not only must we explain stabilization of the electroweak scale against quantum corrections,
but we must also explain why the electroweak scale and TeV-sized SUSY breaking
soft-terms are ``initially'' positioned so far below the Planck mass.
These latter components of the ``gauge hierarchy'' problem are the more subtle.
In their theoretical pursuit, we do not though feign ignorance of
established experimental boundaries, taking the phenomenologist's
perspective that pieces fit already to the puzzle stipulate a partial contour
of those yet to be placed.  Indeed, careful knowledge of
precision electroweak scale physics, including the strong and
electromagnetic couplings, the Weinberg angle and the Z-mass
are required even to run the one loop RGEs.  In the second
loop one requires also {\it minimally} the leading top quark Yukawa coupling, as
deduced from $m_{\rm t}$, and the Higgs VEV
$v \equiv \sqrt{ \langle H_u \rangle ^2 + \langle H_d \rangle ^2} \simeq 174~$GeV,
established in turn from measurement of the effective Fermi coupling,
or from $M_{\rm Z}$ and the electroweak couplings.

Reading the RGEs up from $M_{\rm Z}$,
we take unification of the gauge couplings as evidence of a GUT.
Reading them in reverse from a point of high energy unification, we take the
heaviness of the top quark, via its large Yukawa coupling, to dynamically
drive the term $M_{\rm H_u}^2 +\mu^2$ negative, triggering spontaneous
collapse of the tachyonic vacuum, {\it i.e.}~radiative electroweak symmetry breaking.
Minimization of this potential with respect to the neutral components
of $H_u$ and $H_d$ yields two conditions, which may be solved for
$\mu(M_{\rm Z})$ and $B_\mu(M_{\rm Z})$ in terms of the constrained Higgs VEVs,
which are in turn functions of $M_{\rm Z}$ (considered experimentally fixed) and
$\tan \beta \equiv  \langle H_u \rangle /  \langle H_d \rangle $ (considered a free parameter).

Restricting to just the solution subset for which $B_\mu(M_{\rm Z})$ given by EWSB
stitches cleanly onto that run down under the RGEs from $B_\mu(M_{\cal F})=0$,
$\tan \beta$, or equivalently $\mu$, becomes an implicit
function of the single moduli $M_{1/2}(M_{\cal F})$.
The pinnacle of this construction is the Super No-Scale condition
$d V_{\rm min} / d M_{1/2} = 0$, wherein $M_{1/2}$, and thus also $\tan \beta$,
are dynamically established at the local {\it minimum minimorum}.  By comparison, the standard
MSSM construction seems a hoax, requiring horrendous fine
tuning to stabilize if viewed as a low energy supergravity limit, and moreover
achieving TeV scale EW and SUSY physics as a simple shell game by manual 
selection of TeV scale boundaries for the soft terms $M_{1/2}$, $M_0$, and $A$.

Strictly speaking, having effectively exchanged input of the Z-mass for a constraint
on $\mu (M_{\cal F})$, we dynamically establish the SUSY breaking soft term
$M_{1/2}$ and $\tan \beta$ {\it within} the electroweak symmetry breaking
vacua, {\it i.e.}~with fixed $v\simeq 174$~GeV.
However, having predicted $M_{\cal F}$ as an output scale near
the reduced Planck mass, we are licensed to invert the solution,
taking $M_{\cal F}$ as a high scale {\it input} and dynamically
addressing the gauge hierarchy through the standard story of radiative electroweak
symmetry breaking.  This proximity to the elemental high scale of
(consistently decoupled) gravitational physics, arises because of the
dual flipped unification and the perturbing effect of the TeV multiplets, 
and is not motivated in standard GUTs.  Operating the machinery of the RGEs in reverse,
we may transmute the low scale $M_{\rm Z}$ for the high scale $M_{\cal F}$,
emphasizing that the fundamental dynamic {\it correlation} is that of 
the {\it ratio} $M_{\rm Z}/M_{\cal F}$, taking either as our input yardstick according to taste.
For fixed $M_{\cal F} \simeq 7\times10^{17}$~GeV, in a single breath we receive
the order of the electroweak scale, the Z-mass, the Higgs bilinear coupling $\mu$, the
Higgs VEVs, and all other dependent dimensional quantities, including predictions
for the full superparticle mass spectrum.
It is in this sense that we claim a complete resolution of the gauge hierarchy
problem, within the context of the Super No-Scale $\cal{F}$-$SU(5)$ model.


\section{ The GUT Higgs Modulus }

An alternate pair of parameters for which one may attempt to isolate a $B_{\mu} = 0$ curve,
which we consider for the first time in this work, is that of $M_{1/2}$ and the GUT scale $M_{32}$,
at which the $SU(3)_C$ and $SU(2)_L$ couplings initially meet.  Fundamentally, the latter corresponds
to the modulus which sets the total magnitude of the GUT Higgs field's VEVs.
$M_{32}$ could of course in some sense be considered a ``known'' quantity, taking the low energy couplings as input.
Indeed, starting from the measured SM gauge couplings and fermion Yukawa couplings at the standard 
$91.187$~GeV electroweak scale, we may calculate both $M_{32}$ and the final unification scale $M_{\cal F}$,
and subsequently the unified gauge coupling and SM fermion Yukawa couplings at $M_{\cal F}$, via
running of the RGEs.  However, since the VEVs of the GUT Higgs fields $H$ and $\overline{H}$ are considered
here as free parameters, the GUT scale $M_{32}$ must not be fixed either. As a consequence,
the low energy SM gauge couplings, and in particular the $SU(2)_L$ gauge coupling $g_2$,
will also run freely via this feedback from $M_{32}$.

We consider this conceptual release of a known quantity, in order to establish the
nature of the model's dependence upon it, to be a valid and valuable technique, and have employed it previously
with specific regards to ``postdiction'' of the top quark mass value~\cite{Li:2010mi}.  Indeed, forcing the
theoretical {\it output} of such a parameter is only possible in a model with highly constrained physics, and it
may be expected to meet success only by intervention of either grand coincidence or grand conspiracy of Nature.

For this study, we choose a vector-like particle mass $M_V=1000$~GeV, and use the experimental top
quark mass input $m_{t}=173.1$~GeV.  We emphasize that the choice of $M_V=1000~{\rm GeV}$ is not an arbitrary
one, since a prior analysis~\cite{Li:2010mi} has shown that a $1$~TeV vector-like mass is in compliance
with all current experimental data and the No-Scale $B_{\mu}$=0 requirement.

\begin{figure}[ht]
        \centering
        \includegraphics[width=0.4\textwidth]{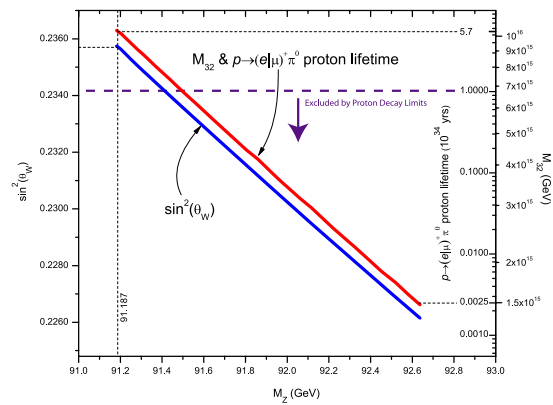}
        \caption{The interrelated variation of $\sin^2 (\theta_{\rm W})$, the GUT scale $M_{32}$ (logarithmic axis),
and the $Z$-boson mass $M_Z$ is demonstrated for the parameter strips which preserve $B_{\mu}=0$ and $\mu=460$~GeV at $M_{\cal F}$.
The variation in $M_Z$ is linked dominantly to motion of the EW couplings via $\sin^2 (\theta_{\rm W})$.
Also shown is the corresponding predicted proton lifetime in the leading
${(e|\mu)}^{+} \pi^0 $ channels, in units of $10^{34}$ years, with the current lower bound of $1.0 \times 10^{34}$ years indicated
by the dashed horizontal purple line.}
\label{fig:sin2T_MZ_M32}
\end{figure}

In actual practice, the variation of $M_{32}$ is achieved in the reverse by programmatic variation of the
Weinberg angle, holding the strong and electromagnetic couplings at their physically measured values.
Figure~\ref{fig:sin2T_MZ_M32} demonstrates the scaling between $\sin^2 (\theta_{\rm W})$,
$M_{32}$ (logarithmic axis), and the $Z$-boson mass.  The variation of $M_Z$ is attributed primarily to the motion
of the electroweak couplings, the magnitude of the Higgs VEV being held essentially constant.
We ensure also that the unified gauge coupling, SM fermion Yukawa couplings, and specifically also
the Higgs bilinear term $\mu \simeq 460$~GeV, are each held stable at the scale $M_{\cal F}$ to correctly
mimic the previously described procedure.

\begin{figure}[ht]
	\centering
	\includegraphics[width=0.4\textwidth]{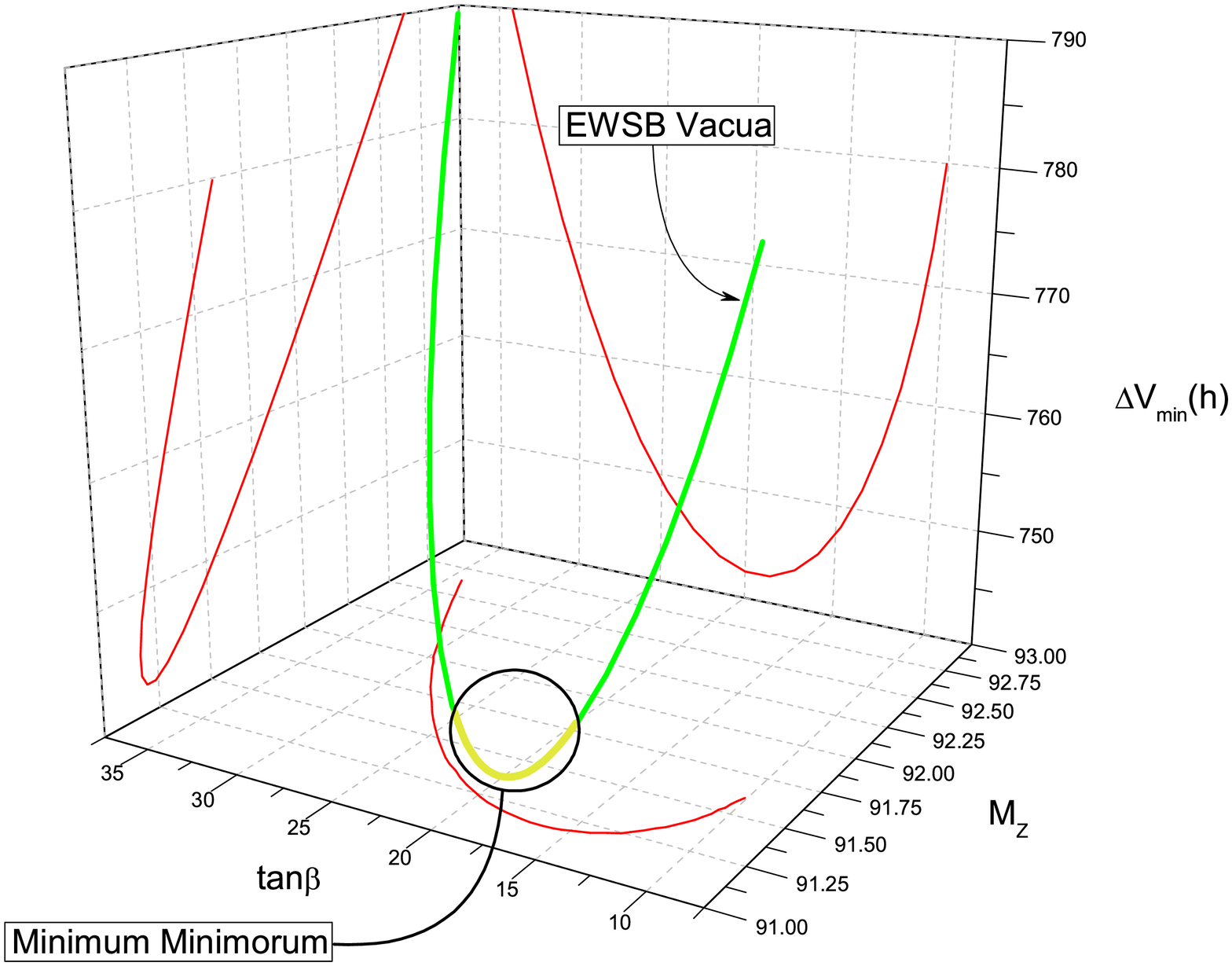}
	\caption{Three-dimensional graph of $(M_{Z},\tan \beta,\Delta V_{min}(h)$) space (green curve). The projections onto the three mutually perpendicular planes (red curves) are likewise shown. $M_{Z}$ and $\Delta V_{min}(h)$ are in units of GeV.  The dynamically preferred region, allowing for plausible variation, is circled and tipped in gold.}
\label{fig:dV_MZ_tanb}
\end{figure}

\begin{figure}[ht]
	\centering
	\includegraphics[width=0.4\textwidth]{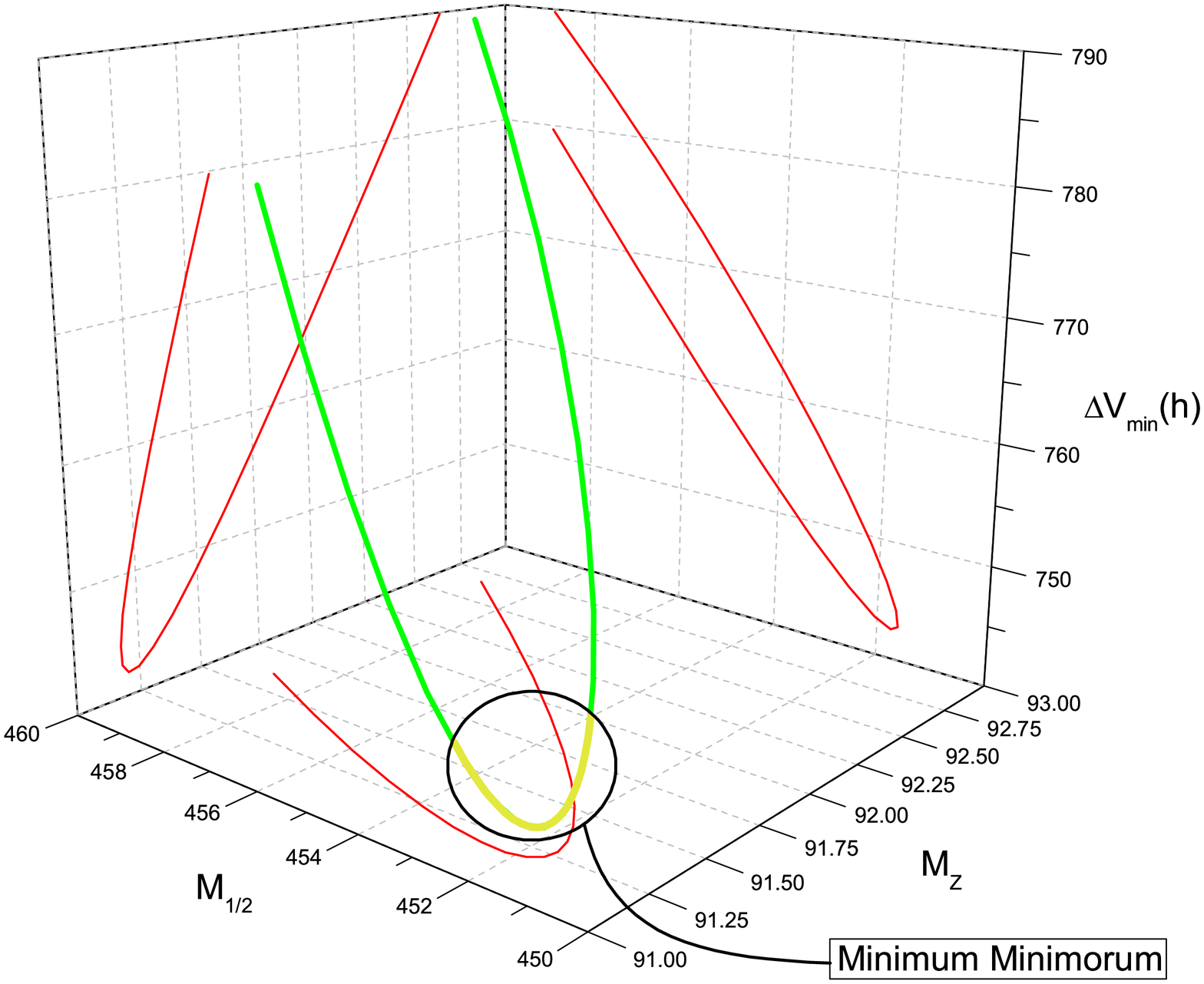}
	\caption{Three-dimensional graph of $(M_{Z},M_{1/2},\Delta V_{min}(h))$ space (green curve). The projections onto the three mutually perpendicular planes (red curves) are likewise shown. $M_{Z}$, $M_{1/2}$, and $\Delta V_{min}(h)$ are in units of GeV.  The dynamically preferred region, allowing for plausible variation, is circled and tipped in gold.}
\label{fig:dV_M12_MZ}
\end{figure}

\begin{figure}[ht]
	\centering
	\includegraphics[width=0.4\textwidth]{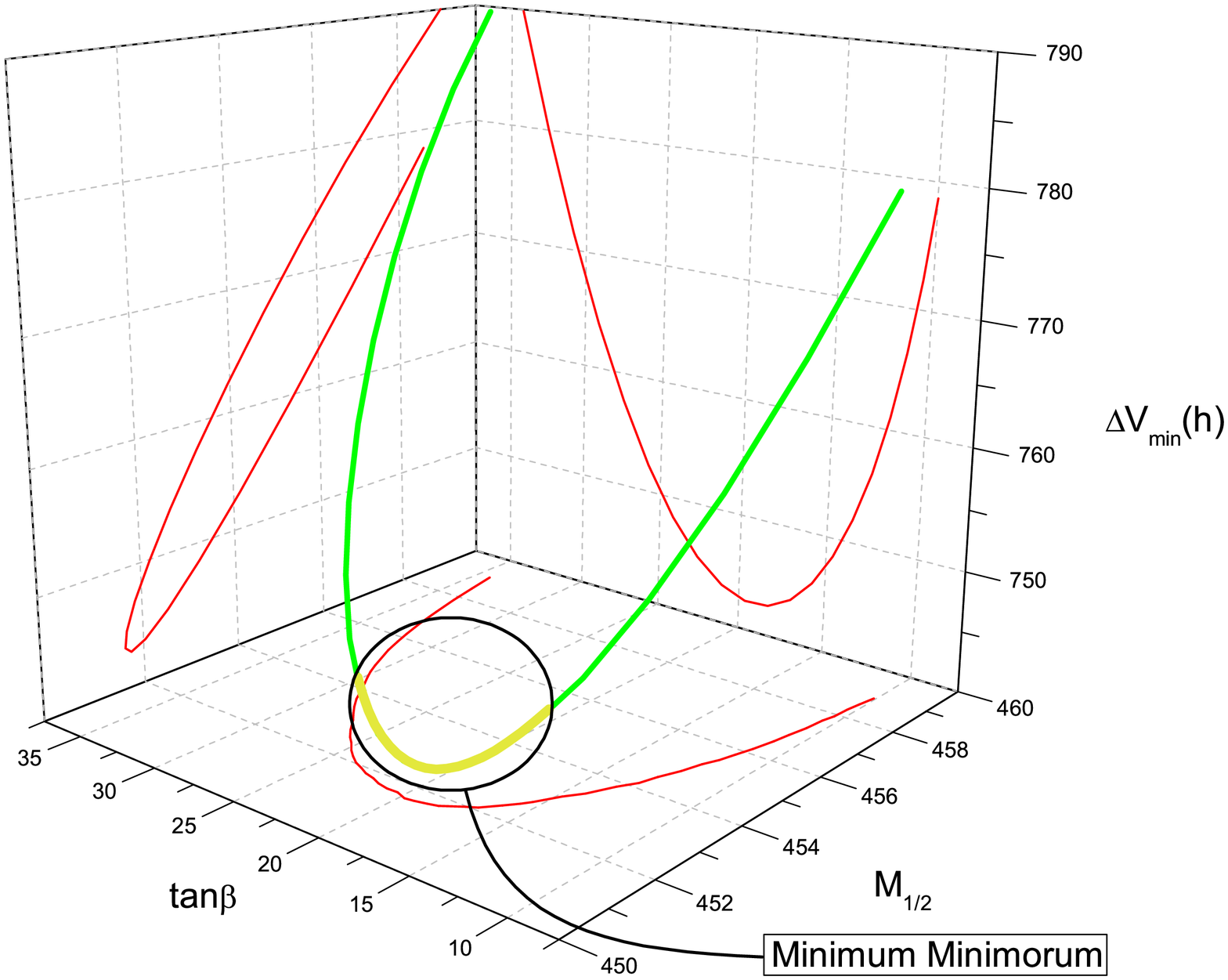}
	\caption{Three-dimensional graph of $(M_{1/2},\tan \beta,\Delta V_{min}(h)$) space (green curve). The projections onto the three mutually perpendicular planes (red curves) are likewise shown. $M_{1/2}$ and $\Delta V_{min}(h)$ are in units of GeV.  The dynamically preferred region, allowing for plausible variation, is circled and tipped in gold.}
\label{fig:dV_M12_tanb}
\end{figure}

\begin{figure}[ht]
	\centering
	\includegraphics[width=0.4\textwidth]{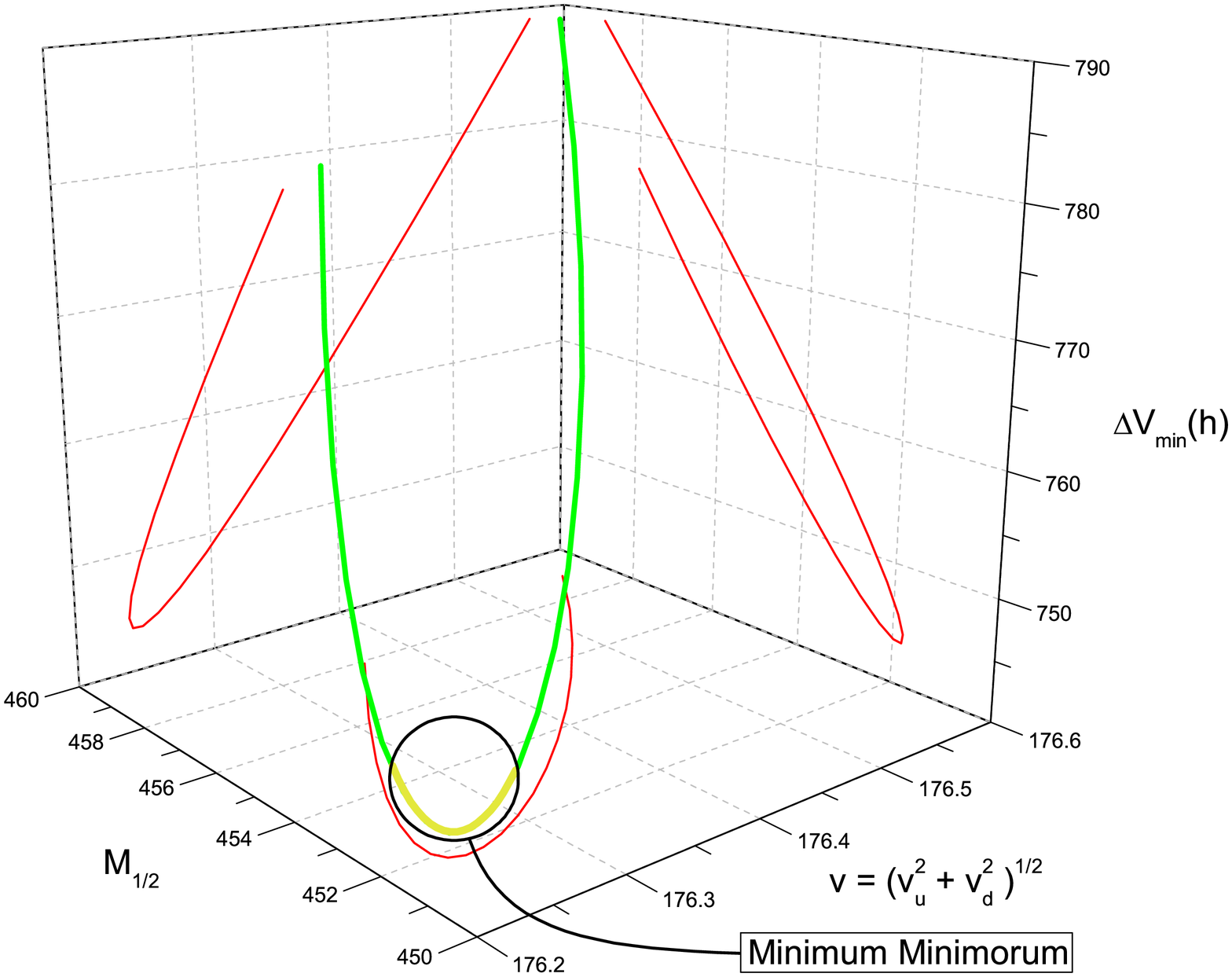}
	\caption{Three-dimensional graph of $(v,M_{1/2},\Delta V_{min}(h))$ space (green curve). The projections onto the three mutually perpendicular planes (red curves) are likewise shown. $M_{1/2}$, $v$, and $\Delta V_{min}(h)$ are in units of GeV.  The dynamically preferred region, allowing for plausible variation, is circled and tipped in gold.}
\label{fig:dV_M12_v}
\end{figure}

\begin{figure}[ht]
	\centering
	\includegraphics[width=0.4\textwidth]{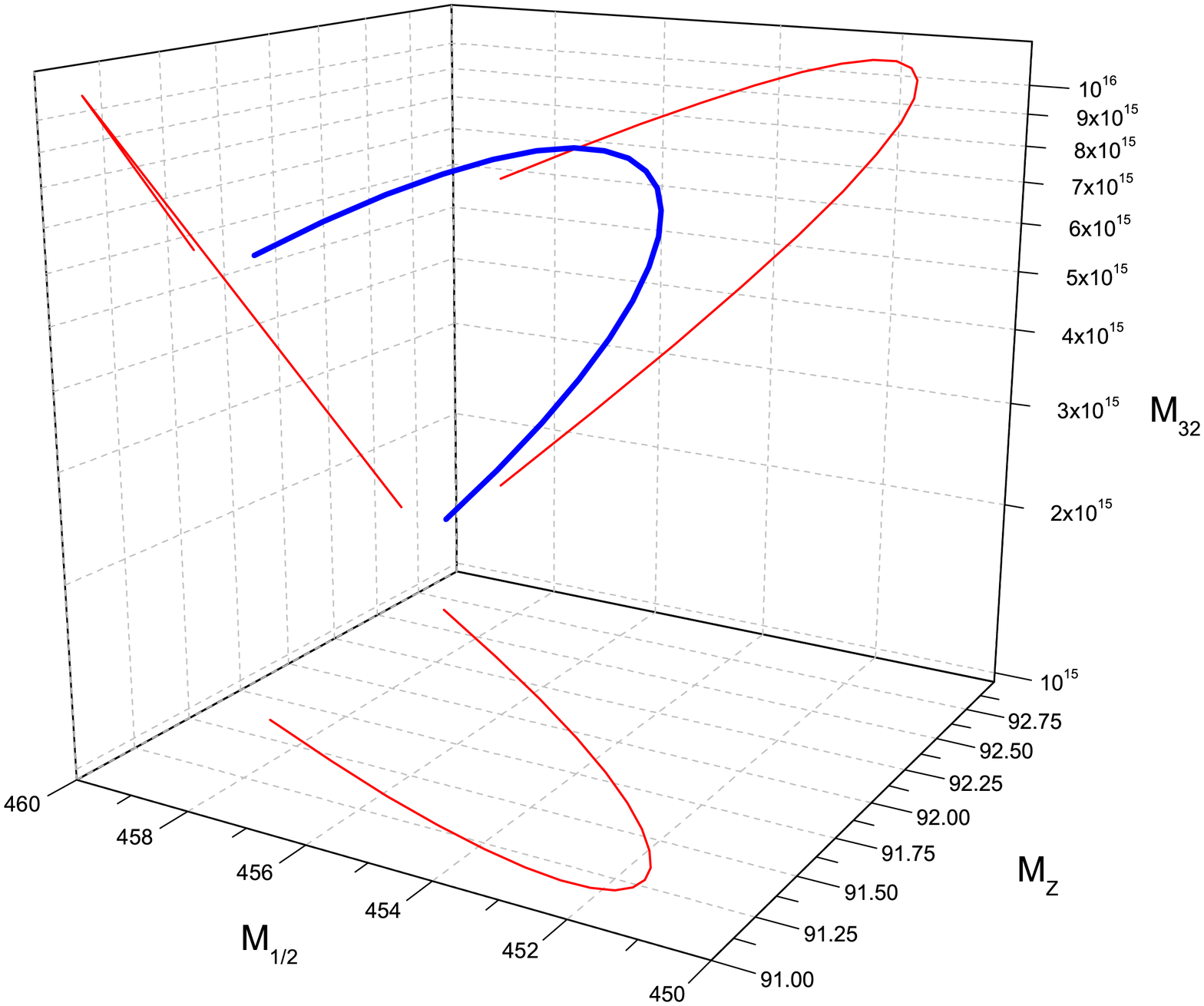}
	\caption{Three-dimensional graph of $(M_{Z},M_{1/2},M_{32})$ space (blue curve). The projections onto the three mutually perpendicular planes (red curves) are likewise shown. $M_{Z}$, $M_{1/2}$, and $M_{32}$ are in units of GeV.}
\label{fig:M32_M12_MZ}
\end{figure}

\begin{figure}[ht]
	\centering
	\includegraphics[width=0.4\textwidth]{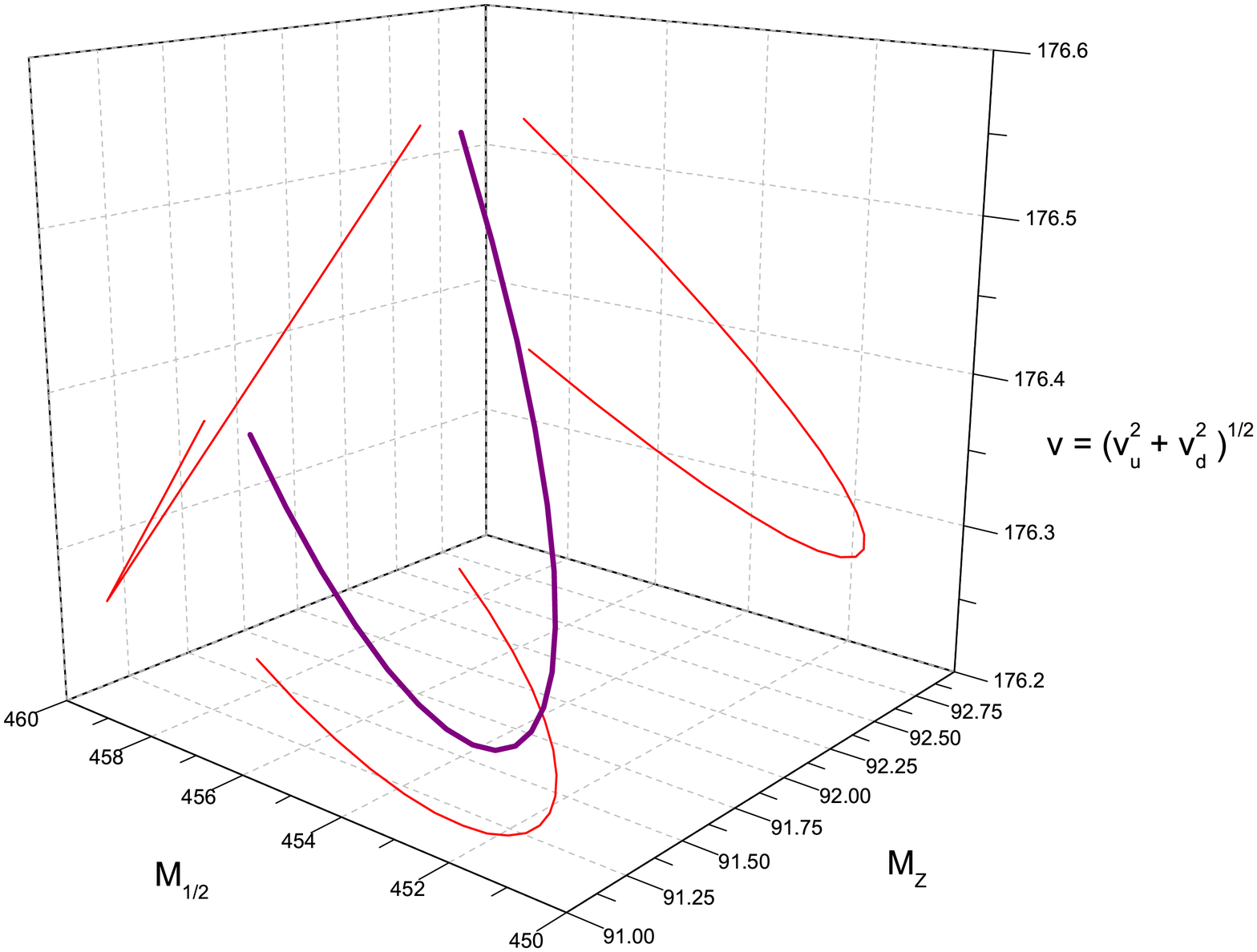}
	\caption{Three-dimensional graph of $(M_{Z},M_{1/2},v)$ space (purple curve). The projections onto the three mutually perpendicular planes (red curves) are likewise shown. $M_{Z}$, $M_{1/2}$, and $v$ are in units of GeV.}
\label{fig:v_M12_MZ}
\end{figure}

\begin{figure}[ht]
	\centering
	\includegraphics[width=0.4\textwidth]{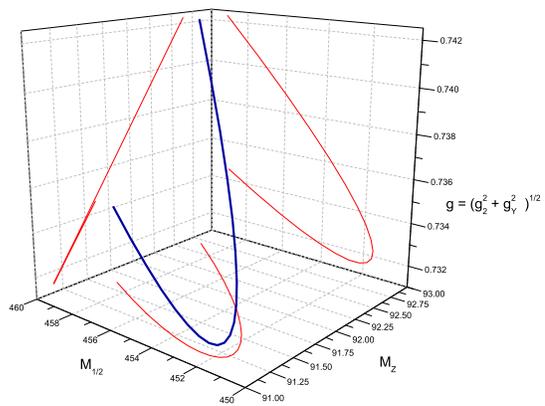}
	\caption{Three-dimensional graph of $(M_{Z},M_{1/2},g)$ space (royal blue curve). The projections onto the three mutually perpendicular planes (red curves) are likewise shown. $M_{Z}$ and $M_{1/2}$ are in units of GeV.}
\label{fig:g_M12_MZ}
\end{figure}

The parameter ranges for the variation depicted in Figure~\ref{fig:sin2T_MZ_M32} are $M_Z = 91.18 - 92.64$,
$\sin^2(\theta_{\rm W}) = 0.2262 - 0.2357$, and $M_{32} = 1.5\times 10^{15} - 1.04 \times 10^{16}$~GeV, and likewise
also the same for Figures~(\ref{fig:dV_MZ_tanb}-\ref{fig:g_M12_MZ}),
which will feature subsequently.  The {\it minimum minimorum} falls at the boundary of the prior list, dynamically
fixing $M_{32} \simeq 1.0 \times 10^{16}$~GeV and placing $M_{1/2}$ again in the vicinity of $450$~GeV.
The low energy SM gauge couplings are simultaneously constrained by means of the associated Weinberg angle, with
$\sin^2 (\theta_{\rm W}) \simeq 0.236$, in excellent agreement with experiment.
The corresponding range of predicted proton lifetimes in the leading ${(e|\mu)}^{+} \pi^0 $ modes is
$2.5\times 10^{31} - 5.7\times10^{34}$~years~\cite{Li:2010dp, Li:2010rz}.  If the GUT scale $M_{32}$ becomes excessively light, below about
$7 \times 10^{15}$~GeV, then proton decay would be more rapid than allowed by 
the recently updated lower bound of $1.0 \times 10^{34}$~years from Super-Kamiokande~\cite{:2009gd}.

We are cautious against making a claim in precisely the same vein for the dynamic determination of $M_Z \simeq 91.2$~GeV, since again
the crucial electroweak Higgs VEV is not a substantial element of the variation.  However, in {\it conjunction} with the radiative electroweak
symmetry breaking~\cite{Ellis:1982wr,Ellis:1983bp} numerically implemented within the {\tt SuSpect 2.34} code base~\cite{Djouadi:2002ze},
the fixing of the Higgs VEV and the determination of the electroweak scale may also plausibly be considered 
legitimate dynamic output, {\it if} one posits the $M_{F}$ scale input to be available {\it a priori}.

The present minimization, referencing $M_{1/2}$, $M_{32}$ and $\tan \beta$, is again dependent upon $M_{V}$ and
$m_{t}$, while the previously described~\cite{Li:2010uu} determination of $\tan \beta$ was, by contrast, $M_{V}$ and $m_{t}$ invariant.
Recognizing that a minimization with all three parameters simultaneously active is required to declare all three
parameters to have been simultaneously dynamically determined, we emphasize the mutual consistency of the results.
We again stress that the new {\it minimum minimorum} is also consistent with the previously advertised
Golden Strip, satisfying all presently known experimental constraints to our available resolution.
It moreover also addresses the problems of the SUSY breaking scale and gauge hierarchy~\cite{Li:2010uu},
insomuch as $M_{1/2}$ is determined dynamically.

\section{ The Minimum Minimorum of the Higgs Potential: Fixing Yukawa Couplings and $\mu$ at $M_{\cal F}$ }

We have revised the {\tt SuSpect 2.34} 
code base~\cite{Djouadi:2002ze} to incorporate our specialized
No-Scale ${\cal F}$-$SU(5)$ with vector-like mass algorithm, and accordingly employ
two-loop RGE running for the SM gauge couplings, and
one-loop RGE running for the SM fermion Yukawa couplings,
$\mu$ term, and SUSY breaking soft terms. For our choice of
$M_V=1000$ GeV, $m_t=173.1$ GeV, and $\mu(M_{\cal F}) \simeq 460$ GeV, 
we present the one-loop effective
Higgs potential $\Delta V_{min}(h)$ in terms of $M_{Z}$ and $\tan \beta$ in
Figure~\ref{fig:dV_MZ_tanb}, in terms of 
$M_{Z}$ and $M_{1/2}$ in
Figure~\ref{fig:dV_M12_MZ},
in terms of 
$M_{1/2}$ and tan$\beta$ in
Figure~\ref{fig:dV_M12_tanb},
and in terms of $v$ and $M_{1/2}$ in
Figure~\ref{fig:dV_M12_v}, where $v=\sqrt{v_{u}^{2}+v_{d}^{2}}$, 
$v_u=\langle H^0_u \rangle$, and $v_d=\langle H^0_d \rangle$.
These figures clearly demonstrate the localization
of the {\it minimum minimorum} of the Higgs potential,
corroborating the dynamical determination of $\tan \beta \simeq 15-20$
and $M_{1/2} \simeq 450$~GeV in~\cite{Li:2010uu}.

Additionally, we exhibit the $(M_{Z},M_{1/2},M_{32})$ space in Figure~\ref{fig:M32_M12_MZ},
the $(M_{Z},M_{1/2},v)$ space in Figure~\ref{fig:v_M12_MZ},
and the $(M_{Z},M_{1/2},g)$ space in Figure~\ref{fig:g_M12_MZ},
where $g=\sqrt{g_{2}^{2}+g_{Y}^{2}}$. Figure~\ref{fig:M32_M12_MZ} demonstrates that $M_{32} \simeq 1.0 \times 10^{16}$~GeV
at the {\it minimum minimorum}, which correlates to $M_{Z} \simeq 91.2$~GeV,
or more directly, $\sin^2 (\theta_{\rm W}) \simeq 0.236$.
Together, the alternate perspectives of Figures~\ref{fig:M32_M12_MZ},
~\ref{fig:v_M12_MZ},
and~\ref{fig:g_M12_MZ} complete the view given in Figures~\ref{fig:dV_MZ_tanb},~\ref{fig:dV_M12_MZ},~\ref{fig:dV_M12_tanb},
and~\ref{fig:dV_M12_v}
to visually tell the story of the dynamic interrelation between the $M_Z$, $M_{1/2}$, and $M_{32}$ scales,
as well as the electroweak gauge couplings, and the Higgs VEVs.
The curves in each of these figures represent only those points that satisfy the $B_{\mu}$ = 0 requirement, as dictated by No-Scale
supergravity, serving as a crucial constraint on the dynamically determined parameter space. Ultimately, it is the significance of
the $B_{\mu} = 0$ requirement that separates the No-Scale ${\cal F}$-$SU(5)$ with vector-like particles from the entire
compilation of prospective string theory derived models. By means of the $B_{\mu}$ = 0 vehicle, No-Scale ${\cal 
 F}$-$SU(5)$ has surmounted the paramount challenge of phenomenology, that of dynamically determining the electroweak scale,
the scale of fundamental prominence in particle physics.

We wish to note that recent progress has been made in incorporating more precise numerical calculations into
our baseline algorithm for No-Scale ${\cal F}$-$SU(5)$ with vector-like particles. Initially, when we commenced the task of fully
developing the phenomenology of this model, the extreme complexity of properly numerically implementing No-Scale
${\cal F}$-$SU(5)$ with vector-like particles compelled a gradual strategy for construction and persistent enhancement of the
algorithm. Preliminary findings of a precision improved algorithm indicate that compliance with the 7-year WMAP relic density
constraints requires a slight upward shift to $\tan \beta \simeq 19-20$ from the value computed in Ref.~\cite{Li:2010ws},
suggesting a potential convergence to even finer resolution of the dynamical determination of $\tan \beta$ given by the Super No-Scale
mechanism, and the value demanded by the experimental relic density measurements.  We shall furnish a comprehensive analysis of the
precision improved algorithm at a later date.


\section{ Probing The Blueprints of the No-Scale Multiverse at the Colliders }

We offer here a brief summary of direct collider, detector, and telescope level tests
which may probe the blueprints of the No-Scale Multiverse which we have laid out.  As to the deep
question of whether the ensemble be literal in manifestation, or merely the conceptual
superset of unrealized possibilities of a single island Universe, we pretend no definitive
answer.  However, we have argued that the emergence {\it ex nihilo} of seedling universes which fuel
an eternal chaotic inflation scenario is particularly plausible, and even natural, within No-Scale
Supergravity, and our goal of probing the specific features of our own Universe which might
implicate its origins in this construction are immediately realizable and practicable.

The unified gaugino $M_{1/2}$ at the unification scale $M_{\cal F}$ can be reconstructed
from impending LHC events by determining the gauginos $M_{1}$, $M_{2}$, and $M_{3}$ at the
electroweak scale, which will in turn require knowledge of the masses for the neutralinos, charginos,
and the gluino. Likewise, $\tan \beta$ can be ascertained in principle from a distinctive experimental
observable, as was accomplished for mSUGRA in~\cite{Arnowitt:2008bz}.  We will not undertake a comprehensive
analysis here of the reconstruction of $M_{1/2}$ and $\tan \beta$, but will offer for now a cursory examination
of typical events expected at the LHC. We will present a detailed compilation of the experimental
observables necessary for validation of the No-Scale ${\cal F}$-$SU(5)$ at the LHC in the following section. 

For the benchmark SUSY spectrum presented in Table~\ref{tab:masses_3}, we have adopted
the specific values $M_{1/2}=453$, $\tan \beta=15$ and $M_Z=91.187$.
We expect that higher order corrections will shift the precise location of the {\it minimum minimorum}
a little bit, for example, within the encircled gold-tipped regions of the diagrams in the prior section.
We have selected a ratio for $\tan \beta$ at the lower end of this range for consistency with our previous
study~\cite{Li:2010uu}, and to avoid stau dark matter.

\begin{table}[ht]
  \small
	\centering
	\caption{Spectrum (in GeV) for the benchmark point. 
Here, $M_{1/2}$ = 453 GeV, $M_{V}$ = 1000 GeV, $m_{t}$ = 173.1 GeV, $M_{Z}$ = 91.187 GeV, $\mu (M_{\cal F})$ = 460.3 GeV, $\Delta V_{min}(h)$ = 748 GeV, $\Omega_{\chi}$ = 0.113, $\sigma_{SI} = 2 \times 10^{-10}$ pb, and $\left\langle \sigma v \right\rangle_{\gamma\gamma} = 1.8 \times 10^{-28} ~cm^{3}/s$. The central prediction for the $p \!\rightarrow\! {(e\vert\mu)}^{\!+}\! \pi^0$ proton lifetime is around $4.9 \times 10^{34}$ years. The lightest neutralino is 99.8\% Bino.}

		\begin{tabular}{|c|c||c|c||c|c||c|c||c|c||c|c|} \hline		
    $\widetilde{\chi}_{1}^{0}$&$94$&$\widetilde{\chi}_{1}^{\pm}$&$184$&$\widetilde{e}_{R}$&$150$&$\widetilde{t}_{1}$&$486$&$\widetilde{u}_{R}$&$947$&$m_{h}$&$120.1$\\ \hline
    $\widetilde{\chi}_{2}^{0}$&$184$&$\widetilde{\chi}_{2}^{\pm}$&$822$&$\widetilde{e}_{L}$&$504$&$\widetilde{t}_{2}$&$906$&$\widetilde{u}_{L}$&$1032$&$m_{A,H}$&$916$\\ \hline
     $\widetilde{\chi}_{3}^{0}$&$817$&$\widetilde{\nu}_{e/\mu}$&$498$&$\widetilde{\tau}_{1}$&$104$&$\widetilde{b}_{1}$&$855$&$\widetilde{d}_{R}$&$988$&$m_{H^{\pm}}$&$921$\\ \hline
    $\widetilde{\chi}_{4}^{0}$&$821$&$\widetilde{\nu}_{\tau}$&$491$&$\widetilde{\tau}_{2}$&$499$&$\widetilde{b}_{2}$&$963$&$\widetilde{d}_{L}$&$1035$&$\widetilde{g}$&$617$\\ \hline
		\end{tabular}
		\label{tab:masses_3}
\end{table}

At the benchmark point, we calculate
$\Omega_{\chi} = 0.113$ for the cold dark matter relic density. 
The phenomenology is moreover consistent with the LEP limit on the lightest CP-even Higgs boson
mass, $m_{h} \geq 114$ GeV~\cite{Barate:2003sz,Yao:2006px}, the CDMS~II~\cite{Ahmed:2008eu} and
Xenon~100~\cite{Aprile:2010um} upper limits on the spin-independent cross section
$\sigma_{SI}$, and the Fermi-LAT space telescope constraints~\cite{Abdo:2010dk} on the 
photon-photon annihilation cross section $\left\langle \sigma v \right\rangle_{\gamma\gamma}$.
The differential cross-sections and branching ratios have been calculated with
{\tt PGS4}~\cite{PGS4} executing a call to {\tt PYTHIA 6.411}~\cite{Sjostrand:2006za}, using our
specialized No-Scale algorithm integrated into the {\tt SuSpect 2.34} code for initial computation of
the sparticle masses.

The benchmark point resides in the region of the experimentally allowed parameter
space that generates the relic density through
stau-neutralino coannihilation. Hence, the five lightest sparticles for this benchmark point are
$\widetilde{\chi}_{1}^{0} < \widetilde{\tau}_{1}^{\pm} < \widetilde{e}_{R} < \widetilde{\chi}_{2}^{0}
\sim \widetilde{\chi}_{1}^{\pm}$. Here, the gluino is lighter than all the squarks with the exception
of the lighter stop, so that all squarks will predominantly decay to a gluino and hadronic jet,
with a small percentage of squarks producing a jet and either a $\widetilde{\chi}_{1}^{\pm}$ or
$\widetilde{\chi}_{2}^{0}$. The gluinos will decay via virtual (off-shell) squarks to neutralinos or charginos
plus quarks, which will further cascade in their decay. The result is a
low-energy tau through the processes $\widetilde{\chi}_{2}^{0} \rightarrow \widetilde{\tau}_{1}^{\mp}
\tau^{\pm} \rightarrow \tau^{\mp}\tau^{\pm} \widetilde{\chi}_{1}^{0}$ and $\widetilde{\chi}_{1}^{\pm}
\rightarrow \widetilde{\tau}_{1}^{\pm} \nu_{\tau} \rightarrow \tau^{\pm}\nu_{\tau} \widetilde{\chi}_{1}^{0}$.

The LHC final states of low-energy tau in the ${\cal F}$-$SU(5)$ stau-neutralino coannihilation region
are similar to those same low-energy LHC final states in mSUGRA, however, in the stau-neutralino
coannihilation region of mSUGRA, the gluino is typically heavier than the squarks.
The LHC final low-energy
tau states in the stau-neutralino coannihilation regions of ${\cal F}$-$SU(5)$ and mSUGRA will thus differ
in that in ${\cal F}$-$SU(5)$, the low-energy tau states will result largely from neutralinos and
charginos produced by gluinos, as opposed to the low-energy tau states in mSUGRA resulting primarily
from neutralinos and charginos produced from squarks.

Also notably, the TeV-scale vector-like multiplets are well targeted
for observation by the LHC.  We have argued~\cite{Li:2010mi} that the eminently
feasible near-term detectability of these hypothetical fields in collider experiments,
coupled with the distinctive flipped charge assignments within the multiplet structure,
represents a smoking gun signature for Flipped $SU(5)$, and have thus coined the term
{\it flippons} to collectively describe them.
Immediately, our curiosity is piqued by the recent announcement~\cite{Abazov:2010ku}
of the D\O~collaboration that vector-like quarks have been excluded up to
a bound of 693~GeV, corresponding to the immediate lower edge of our anticipated
range for their discovery~\cite{Li:2010mi}.


\section{The Ultra-High Jet Signal of No-Scale $\cal{F}$-$SU(5)$ at the $\sqrt{s}=7$ TeV LHC}

The Large Hadron Collider (LHC) at CERN has been accumulating 
data from ${\sqrt s}=7$ TeV proton-proton collisions since March 2010. It is expected 
to reach an integrated luminosity of $1~{\rm fb}^{-1}$ by the end of 2011, all 
in search of new physics beyond the SM.  SUSY, which provides a natural 
solution to  the gauge hierarchy problem, is the most promising
 extension of the SM.  Data corresponding to a limited $35~{\rm pb}^{-1}$
has already established new constraints on the viable parameter
space~\cite{Khachatryan:2011tk, daCosta:2011hh, daCosta:2011qk} due to the unprecedented center of mass
collision energy.  The search strategy for SUSY signals in early LHC data has
been actively and eagerly studied by quite a few groups~\cite{Baer:2010tk,Kane:2011zd, Feldman:2011me, Buchmueller:2011aa, Guchait:2011fb},
 with particular focus
on the parameter space featuring a traditional mass relationship between squarks 
and the gluino, such as a gluino heavier than all squarks or a gluino lighter than all squarks.

A question of great interest is whether there exist SUSY models which are well 
motivated by a fundamental theory such as string theory, 
which can be tested in the initial LHC run, permitting 
a probe of the UV physics close to the Planck scale.
In this Section we present 
such a model. It is well known that the supersymmetric flipped 
$SU(5)\times U(1)_X$ models can solve
the doublet-triplet splitting problem elegantly via the missing
partner mechanism~\cite{Antoniadis:1987tv, Antoniadis:1988tt,Antoniadis:1989zy}. To realize the string scale gauge coupling
unification, two of us (TL and DVN) with Jiang proposed the testable 
flipped $SU(5)\times U(1)_X$ models with TeV-scale vector-like 
particles~\cite{Jiang:2006hf}, where such models can be 
realized in the ${\cal F}$-ree ${\cal F}$-ermionic string
constructions~\cite{Lopez:1992kg}
 and ${\cal F}$-theory model building~\cite{Jiang:2009zza,Jiang:2009za}, dubbed $\cal{F}$-$SU(5)$.
In particular, we find the generic phenomenological consequences are quite
interesting~\cite{Jiang:2009zza, Jiang:2009za, Li:2009fq}. 

In the simplest No-Scale supergravity, 
all the SUSY breaking soft terms arise from a single 
parameter $M_{1/2}$. The spectra in the 
entire Golden Strip are therefore very similar up to a small 
rescaling on $M_{1/2}$, with equivalent sparticle branching 
ratios.  This leaves invariant most of the ``internal'' 
physical properties, whereas this rescaling ability on $M_{1/2}$
 is not apparent in alternative SUSY models. 
For our analysis here, we use a vector-like particle mass of $M_V \sim 1000 $ GeV, which exists in the viable
parameter space; We emphasize that this is not an arbitrary choice~\cite{Li:2010mi}, though these vector-like
particles could be to heavy for observation in the early LHC run.
The gluino mass is about 550 GeV, which should by contrast allow for direct
testing of No-Scale $\cal{F}$-$SU(5)$ at the early LHC run. 

To represent our model for this phase of analysis, we select the
No-Scale ${\cal F}$-$SU(5)$
benchmark point of Table~\ref{tab:masses_4}.
The optimized signatures presented here offer an alluring testing 
vehicle for the stringy origin of $\cal{F}$-$SU(5)$.
This point is again representative of the entire highly 
constrained ${\cal F}$-$SU(5)$ viable
parameter space. The SUSY breaking parameters 
for this point slightly differ from previous ${\cal F}$-$SU(5)$ 
studies~\cite{Li:2010ws,Li:2010mi,Li:2011dw} insomuch as more precise 
numerical calculations have been incorporated into our baseline algorithm.
The masses shift a few GeV from the spectra given in previous work, 
but where different, we believe this to be the more accurate representation.
The branching ratios and decay modes of the spectrum in Table~\ref{tab:masses_4}
and the spectra in~\cite{Li:2010ws,Li:2010mi,Li:2011dw} 
are identical, so the physical properties are consistent before and 
after code improvements. Thus, the signatures studied here will be 
common to the spectra provided previously.

\begin{table}[htbp]
  \small
	\centering
	\caption{Spectrum (in GeV) for $M_{1/2}$ = 410 GeV, $M_{V}$ = 1 TeV, $m_{t}$ = 174.2 GeV, tan$\beta$ = 19.5. Here, $\Omega_{\chi}$ = 0.11 and the lightest neutralino is 99.8\% bino.}
		\begin{tabular}{|c|c||c|c||c|c||c|c||c|c||c|c|} \hline		
    $\widetilde{\chi}_{1}^{0}$&$76$&$\widetilde{\chi}_{1}^{\pm}$&$165$&$\widetilde{e}_{R}$&$157$&$\widetilde{t}_{1}$&$423$&$\widetilde{u}_{R}$&$865$&$m_{h}$&$120.4$\\ \hline
    $\widetilde{\chi}_{2}^{0}$&$165$&$\widetilde{\chi}_{2}^{\pm}$&$756$&$\widetilde{e}_{L}$&$469$&$\widetilde{t}_{2}$&$821$&$\widetilde{u}_{L}$&$939$&$m_{A,H}$&$814$\\ \hline
    
    $\widetilde{\chi}_{3}^{0}$&$752$&$\widetilde{\nu}_{e/\mu}$&$462$&$\widetilde{\tau}_{1}$&$85$&$\widetilde{b}_{1}$&$761$&$\widetilde{d}_{R}$&$900$&$m_{H^{\pm}}$&$820$\\ \hline
    $\widetilde{\chi}_{4}^{0}$&$755$&$\widetilde{\nu}_{\tau}$&$452$&$\widetilde{\tau}_{2}$&$462$&$\widetilde{b}_{2}$&$864$&$\widetilde{d}_{L}$&$942$&$\widetilde{g}$&$561$\\ \hline
		\end{tabular}
		\label{tab:masses_4}
\end{table}

For the initial phase of generation of the low order Feynman diagrams which may link the incoming beam to
the desired range of hard scattering intermediate states, we have used the program {\tt MadGraph 4.4}~\cite{Alwall:2007st}. These diagrams were subsequently fed into {\tt MadEvent}~\cite{Alwall:2007st} for appropriate kinematic scaling to yield batches of Monte Carlo simulated parton level scattering events. The cascaded fragmentation and hadronization of these events into
final state showers of photons, leptons, and mixed jets has been handled by {\tt PYTHIA}~\cite{Sjostrand:2006za}, with {\tt PGS4}~\cite{PGS4} simulating the physical detector environment. We implement MLM matching to preclude double counting of final states, and use the CTEQ6L1 parton distribution functions to generate the SM background. All 2-body SUSY processes are simulated. The b-jet tagging algorithm in {\tt PGS4} is adjusted to update the b-tagging efficiency to $\sim$60\%. We veto an event if any of the following conditions are met: $p_{T}$ $<$ 100 GeV for the two leading jets; $p_{T}$ $<$ 350 GeV for all jets; pseudorapidity $|\eta|$ $>$ 2 for the leading jet; missing energy $ {E \! \! \! \! / \hspace{0.02in}}_{T} < 150$~GeV; isolated photon with $p_{T}$ $>$ 25 GeV; or isolated electron or muon with $p_{T}$ $>$ 10 GeV. Likewise, we discard any single jet with $|\eta|$ $>$ 3. These cuts are quite standard, but alone they are insufficient to reveal the ultra-high multiplicity jet event signature; We must also investigate the event cut on the number of jets and the $p_{T}$ cut on a single jet to preserve ultra-high jet events.

The detector simulations use the spectrum for the ${\cal F}$-$SU(5)$ point 
in Table~\ref{tab:masses_4}. The most significant asset of the spectrum for our analysis is the relationship between the stop, gluino, and other squarks. The distinctive mass pattern of $m_{\widetilde{t}_1} < m_{\widetilde{g}} < m_{\widetilde{q}}$ is the smoking gun signature and possibly a unique characteristic of only ${\cal F}$-$SU(5)$. To gain a comparison of the model studied here with more standardized SUSY models, we examine the ten ``Snowmass Points and Slopes'' (SPS) benchmark points~\cite{Allanach:2002nj} for suitable samples. We find that none of the ten SPS benchmarks support the $m_{\widetilde{t}_1} < m_{\widetilde{g}} < m_{\widetilde{q}}$ mass pattern. This critical element is indicative of how unique the ${\cal F}$-$SU(5)$ signal could be. Previous minimal supersymmetric SM studies focused on signals from a low-multiplicity of jets, whereas the aforementioned mass pattern is expected to show a very high-multiplicity of jets. For the SPS benchmarks, we only consider those spectra not light enough to have been excluded by the initial phase of LHC data, or those not too heavy for early LHC production. A few points satisfy these criteria, though we select only one since we anticipate the corollary points to exhibit analogous characteristics. For our analysis here, we use the SPS SP3 benchmark.

\begin{figure*}[htbp]
        \centering
        \includegraphics[width=1.00\textwidth]{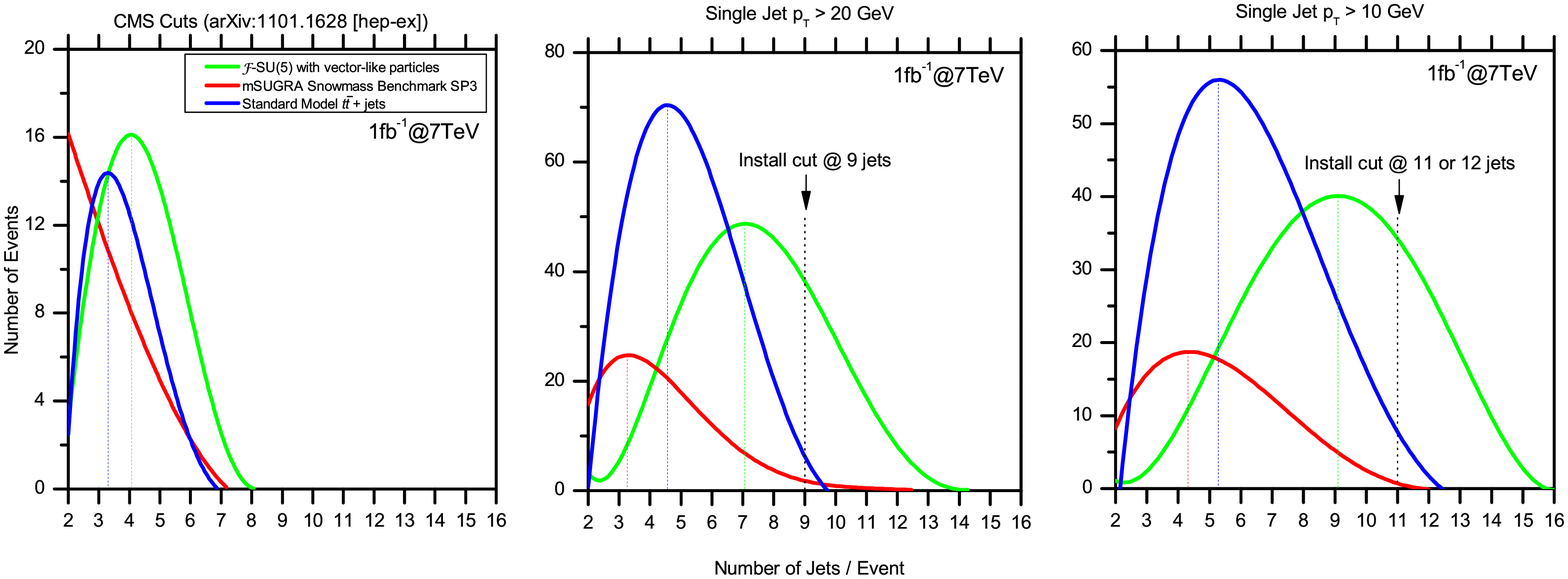}
        \caption{Distribution of events per number of jets. For clarity of the peaks, polynomials have been fitted over the histograms.}
\label{fig:jet_comp}
\end{figure*}

Considering the large number of hadronic jets we are examining for our signatures, there is little intrusion from SM background processes after post-processing cuts. We examine the background processes studied in~\cite{Baer:2010tk,Kane:2011zd, Feldman:2011me, Buchmueller:2011aa, Guchait:2011fb,Altunkaynak:2010we} and our conclusion is that only the $t \overline{t} + jets$ possesses the requisite minimum cross-section and sufficient number of jets to intrude upon the ${\cal F}$-$SU(5)$ signatures. Processes with a higher multiplicity of top quarks can generate events with a large number of jets, however, the cross-sections are sufficiently suppressed to be negligible, bearing in mind the large number of ultra-high jet events which our model will generate. The same is true for those more complicated background processes involving combinations of top quarks, jets, and one or more vector bosons, where the production counts for 1 $fb^{-1}$ of luminosity are again sufficiently small. Furthermore, we neglect the QCD 2,3,4 jets, one or more vector bosons, and $b \overline{b}$ processes since none of these can sufficiently produce events with 9 or more jets after post-processing cuts have been applied.

\begin{figure}[ht]
        \centering
        \includegraphics[width=0.4\textwidth]{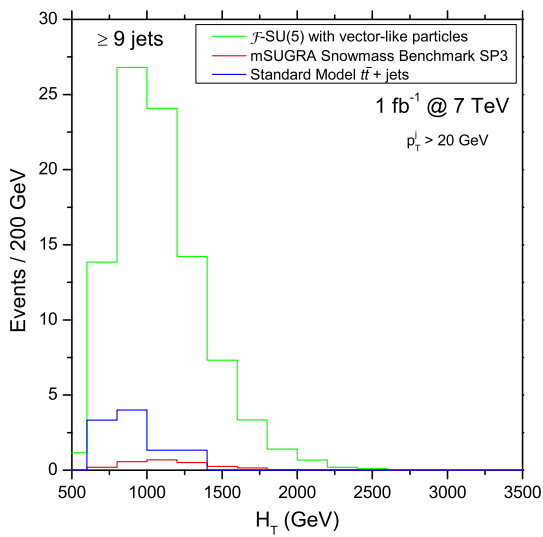}
        \caption{Counts for events with $\ge$ 9 jets.}
\label{fig:gte9jets}
\end{figure}

The ${\cal F}$-$SU(5)$ with vector-like particles mass pattern produces events with a high multiplicity of virtual stops, which concludes in events with a very large number of jets through the dominant chains $\widetilde{g} \rightarrow \widetilde{t}_{1} \overline{t} \rightarrow t \overline{t} \widetilde{\chi}_1^{0} \rightarrow W^{+}W^{-} b \overline{b} \widetilde{\chi}_1^{0}$ and $\widetilde{g} \rightarrow \widetilde{t}_{1} \overline{t} \rightarrow b \overline{t} \widetilde{\chi}_1^{+} \rightarrow W^{-} b \overline{b} \widetilde{\tau}_{1}^{+} \nu_{\tau} \rightarrow W^{-} b \overline{b} \tau^{+} \nu_{\tau} \widetilde{\chi}_1^{0}$, as well as the conjugate processes $\widetilde{g} \rightarrow \widetilde{\overline{t}}_{1} t \rightarrow t \overline{t} \widetilde{\chi}_1^{0}$ and $\widetilde{g} \rightarrow \widetilde{\overline{t}}_{1} t \rightarrow \overline{b} t \widetilde{\chi}_1^{-}$, where the $W$ bosons will produce mostly hadronic jets and some leptons. Additionally, the heavy squarks will produce gluinos by means of $\widetilde{q} \rightarrow q \widetilde{g}$. In Figure~\ref{fig:jet_comp} we plot the number of jets per event versus the number of events for three distinct scenarios. We suppress the noise on the histogram contour to admit a more lucid distinction of the peaks in the number of jets, and fit polynomials over the data points and conceal the histograms. This allows us to gauge an appropriate selection cut for the number of jets to maximize our signal to background ratio, while assessing the impact of the selection cuts implemented by the CMS Collaboration in~\cite{Khachatryan:2011tk,PAS-SUS-09-001}. As depicted in Figure~\ref{fig:jet_comp}, the first pane displays a comparison of the number of jets when employing the prior CMS cuts, while the remaining two panes present the results for the post-processing selection cuts defined in this paper, discriminating between two explicit cuts of the minimum $p_{T}$ for a single jet. Figure~\ref{fig:jet_comp} demonstrates that the CMS cuts of~\cite{Khachatryan:2011tk,PAS-SUS-09-001} discard all the high-multiplicity jets, converting the events with at least 9 jets to events with few jets, thus, all information on these events with a large number of jets is lost. To retain the events with a high multiplicity of jets, we explore alternative cuts by shifting the minimum $p_{T}$ for a single jet lower to the two cases of 10 GeV and 20 GeV. A minimum jet $p_{T}$ of 20 GeV is secure from interfering with jet fragmentation, which typically occurs in the realm below 10 GeV, indicating that 10 GeV is certainly fringe. We see in Figure~\ref{fig:jet_comp} that both the 10 GeV and 20 GeV jet $p_{T}$ cuts preserve the high number of jets, permitting an obvious choice for location of the cut on the minimum number of jets. We thus adopt a revised cut of single jet $p_{T} >$ 20 GeV and total number of jets greater than 9. To assess the discovery potential, we plot the number of events per 200 GeV versus $H_{T}$, where $H_{T} = \sum_{i=1}^{N_{jet}}E_{T}^{j_{i}}$. Figure~\ref{fig:gte9jets} delineates the convincing separation between the ${\cal F}$-$SU(5)$ signal and the SM $t \overline{t} + jets$ and the SP3 point. The total number of events are summarized in Table~\ref{tab:counts}. We also include one measure of discovery threshold that compares the number of signal events S to the number of background events B, where we require $\frac{S}{\sqrt{B}} >$ 5. Notice that ${\cal F}$-$SU(5)$ comfortably surpasses this requirement.

\begin{table}[ht]
  \small
	\centering
	\caption{Total number of events for 1 $fb^{-1}$ and $\sqrt{s}$ = 7 TeV. Minimum $p_{T}$ for a single jet is $p_{T} >$ 20 GeV.}
		\begin{tabular}{|c|c|c|c|} \hline
		$$&${\cal F}$-$SU(5)$&$SP3$&$t\overline{t}+jets$\\ \hline\hline
    $Events$&$93.2$&$2.4$&$10$\\ \hline
    $\frac{S}{\sqrt{B}}$&$29.5$&$0.76$&$$\\ \hline
		\end{tabular}
		\label{tab:counts}
\end{table}

The spectrum of Table~\ref{tab:masses_4} exceeds the LEP constraints on the lightest neutralino $\widetilde{\chi}_{1}^{0}$ and lightest stau $\widetilde{\tau}_{1}$, and even more tantalizing, the close proximity of the stau mass beyond the LEP reach suggests imminent discovery at LHC. The stau presence can be reconstructed, for instance, from the dominant ${\cal F}$-$SU(5)$ process $\widetilde{g} \rightarrow \widetilde{t}_{1} \overline{t} \rightarrow b \overline{t} \widetilde{\chi}_1^{+} \rightarrow W^{-} b \overline{b} \widetilde{\tau}_{1}^{+} \nu_{\tau} \rightarrow W^{-} b \overline{b} \tau^{+} \nu_{\tau} \widetilde{\chi}_1^{0}$. The inference of the short-lived stau in the ${\cal F}$-$SU(5)$ SUSY breaking scenario from tau production assumes fruition of the expected much improved tau detection efficiency at LHC.


\section{Conclusion}

The advancement of human scientific knowledge and technology is replete with instances of science fiction transitioning to scientific
theory and eventually scientific fact. The conceptual notion of a ``Multiverse'' has long fascinated the human imagination,
though this speculation has been largely devoid of a substantive underpinning in physical theory.  The modern perspective presented
here offers a tangible foundation upon which legitimate discussion and theoretical advancement of the Multiverse may commence,
including the prescription of specific experimental tests which could either falsify or enhance the viability of our proposal.
Our perspective diverges from the common appeals to statistics and the anthropic principle, suggesting instead that we may seek
to establish the character of the master theory, of which our Universe is an isolated vacuum condensation, based on specific observed properties
of our own physics which might be reasonably inferred to represent invariant common characteristics of all possible universes.  We have
focused on the discovery of a model universe consonant with our observable phenomenology, presenting it as confirmation of a
non-zero probability of our own Universe transpiring within the larger String Landscape.

The archetype model universe which we advance in this work implicates No-Scale supergravity as
the ubiquitous supporting structure which pervades the vacua of the Multiverse,
being the crucial ingredient in the emanation of a cosmologically flat universe from the quantum ``nothingness''.
In particular, the model dubbed No-Scale ${\cal F}$-$SU(5)$ has demonstrated remarkable consistency between parameters
determined dynamically (the top-down approach) and parameters determined through the application of current
experimental constraints (the bottom-up approach).  This enticing convergence of theory with experiment
elevates No-Scale ${\cal F}$-$SU(5)$, in our estimation, to a position as the current leading GUT candidate.
The longer term viability of this suggestion is likely to be greatly clarified in the next few years, based upon the wealth of
forthcoming experimental data.

We have presented a highly constrained ``Golden Point'' located 
near $M_{1/2} = 455$~GeV and $\tan \beta = 15$ in the $\tan\beta-M_{1/2}$ plane,
and a highly non-trivial ``Golden Strip'' with $\tan\beta\simeq 15$,
$m_{\rm t} = 173.0$-$174.4$~GeV, $M_{1/2} = 455$-$481$~GeV, and $M_{\rm V} = 691$-$1020$~GeV,
which simultaneously satisfies all the known experimental constraints, featuring moreover
an imminently observable proton decay rate. In addition, we have studied the one-loop effective
Higgs potential, and considered the ``Super-No-Scale'' condition.  With a fixed $Z$-boson
mass, we dynamically determined $\tan\beta$ and $M_{1/2}$ at the local
{\it minimum minimorum} of the Higgs potential, while simultaneously indirectly
determining the electroweak scale, thus suggesting 
a complete resolution of the gauge hierarchy problem in the
Standard Model (SM). Furthermore, fixing the SM fermion 
Yukawa couplings and $\mu$ term at
the $SU(5)\times U(1)_X$ unification scale, we 
dynamically determine the ratio $\tan \beta \simeq 15-20$,
the universal gaugino boundary mass $M_{1/2} \simeq 450$~GeV,
and consequently also the total magnitude of the GUT-scale Higgs VEVs,
while constraining the low energy SM gauge couplings.
In particular, these local {\it minima minimorum} lie within the previously described
``Golden Strip'', satisfying all current experimental constraints.

The LHC era has long been anticipated for the expected revelations of physics beyond the Standard Model, as the quest for experimental evidence and insight into the structure of the underlying theory at high energies is enticingly close at hand. Consequently, the field of prospective supersymmetry models has grown as fingerprints of these models at LHC are studied. Nevertheless, our exploration of recently published signatures for supersymmetry discovery reveals a common focus toward low-multiplicity jet events. However, we showed here that manipulation of LHC data skewed toward these low jet events could mask an authentic supersymmetry signal. We have offered a clear and convincing ultra-high jet multiplicity signal for events with at least nine jets, unmistakable for the Standard Model or minimal supergravity. Notably, the optimized post-processing selection cuts outlined here are essential for discovery of supersymmetry if ${\cal F}$-$SU(5)$ is indeed proximal to the physical model. Our revised cuts are not drastic, with the two chief adjustments being lowering the minimum $p_{T}$ for a single jet to 20 GeV, and raising the minimum number of jets in an event to nine. Recognition of such a signal of stringy origin at the LHC could not only reveal the flipped nature of the high-energy theory, but might also shed light on the geometry of the hidden compactified six-dimensional manifold in the string derived models, and even possibly on the hidden structure of the No-Scale Multiverse.

The blueprints which we have outlined here, integrating precision phenomenology with prevailing experimental data and
a fresh interpretation of the Multiverse and the Landscape of String vacua, offer a
logically connected point of view from which additional investigation may be mounted.
As we anticipate the impending stream of new experimental data which is likely to be revealed in ensuing years,
we look forward to serious discussion and investigation of the perspective presented in this work.
Though the mind boggles to contemplate the implications of this speculation, so it must also
reel at even the undisputed realities of the Universe, these acknowledged facts alone being
manifestly sufficient to humble our provincial notions of longevity, extent, and largess.
The stakes could not be higher or the potential revelations more profound.


\section*{Acknowledgments}
D.V.N. would like to thank Thomas Elze for his kind invitation to DICE 2010.
This research was supported in part
by  the DOE grant DE-FG03-95-Er-40917 (TL and DVN),
by the Natural Science Foundation of China
under grant numbers 10821504 and 11075194 (TL),
and by the Mitchell-Heep Chair in High Energy Physics (JM).


\bibliography{bibliography}

\end{document}